\documentclass[prd,nofootinbib,preprintnumbers,preprint,
floatfix]{revtex4}
\usepackage{empheq}

\usepackage{amsmath}
\usepackage{amsfonts}
\usepackage{amssymb}
\usepackage{dsfont}
\usepackage{bm}
\usepackage[hyperfootnotes=false]{hyperref}
\usepackage[dvipsnames]{xcolor}
\usepackage{apptools}
\usepackage{enumerate}

\newcommand{\G}{\mathcal G}
\newcommand{\M}{\mathcal M}

\newcommand{\F}{\mathcal F}

\newcommand{\RS}{\mathcal R}

\newcommand{\LCM}{\nabla}
\newcommand{\LCS}{\mathcal D}
\newcommand{\SFS}{\chi}

\newcounter{example}[section]
\newcounter{remark}[section]
\newcounter{theorem}[section]
\newcounter{proposition}[section]
\newcounter{lemma}[section]
\newcounter{corollary}[section]
\newcounter{definition}[section]

\AtAppendix{\counterwithin{example}{section}}
\AtAppendix{\counterwithin{remark}{section}}
\AtAppendix{\counterwithin{theorem}{section}}
\AtAppendix{\counterwithin{proposition}{section}}
\AtAppendix{\counterwithin{lemma}{section}}
\AtAppendix{\counterwithin{corollary}{section}}
\AtAppendix{\counterwithin{definition}{section}}

\def\theremark{\arabic{section}.\arabic{remark}}
\def\thetheorem{\arabic{section}.\arabic{theorem}}

\def\thedefinition{\arabic{section}.\arabic{definition}}

\renewcommand*{\email}[1]{\footnote{Electronic address: \href{mailto:#1}{\nolinkurl{#1}} }}

\newenvironment{theorem}{\refstepcounter{theorem}
\medskip\noindent{\bf Theorem \thetheorem}:\em}{\medskip}

\begin{document}

\title{Black hole shadows of massive particles and photons in plasma}
\author{Kirill Kobialko${}^{1,\,}$\email{kobyalkokv@yandex.ru}}
\author{Igor Bogush${}^{2,\,}$\email{igbogush@gmail.com}}
\author{Dmitri Gal'tsov${}^{1,\,}$\email{galtsov@phys.msu.ru}}
\affiliation{${}^1$ Faculty of Physics, Moscow State University, 119899, Moscow, Russia\\
${}^2$ Moldova State University, str. A. Mateevici 60, 2009, Chi\c{s}in\u{a}u, Republic of Moldova}

\begin{abstract}
 
Explicitly covariant analytical expressions are derived that describe the boundaries of shadows cast by massive particles scattered by a gravitating object. This covers scenarios with particles having effectively variable mass, such as photons in plasma, geodesics in higher dimensions, and particles interacting with a scalar field. The derived formula takes advantage of recent advances in understanding the relationship between slice-reducible Killing tensors and massive particle surfaces that generalize photon surfaces. The formula allows us to obtain simple approximations of scaling as the particle energy changes. We illustrate this structure using Kerr-NUT and EMD black holes for both massive particles and photons in plasma. The versatility of this framework extends beyond astrophysics and has potential applications in analog models of gravity and condensed matter physics.

\end{abstract}

\maketitle

\setcounter{page}{2}

\setcounter{equation}{0}
\setcounter{subsection}{0}

\section{Introduction}

Spectacular success of the Event Horizon Telescope (EHT) collaboration in constructing shadows of supermassive black holes in M87 and our galaxies \cite{EventHorizonTelescope:2019dse,EventHorizonTelescope:2022wkp} stimulated rapid development in the theoretical description of strong gravitation lensing in a closed vicinity of the event horizon (for review see \cite{Perlick:2021aok,Bronzwaer:2021lzo,Dokuchaev:2020wqk,Dokuchaev:2019jqq,Cunha:2018acu,Ishkaeva:2023xny}). It was soon realised that crucial role in understanding of the observed pictures is played by the {\it photon surfaces} \cite{Claudel:2000yi} -- compact surfaces outside the event horizon where the compact photon orbits are located. It turned out that apart of the picture of these surfaces as filled by compact null geodesics they can be usefully presented as three-dimensional submanifolds in spacetime satisfying the {\it ubmilicity} condition: proportionality of the induced metric and the extrinsic curvature tensor \cite{Chen}. This purely geometric property can serve as a constructive definition of photon surfaces instead of using geodesic equations. Photon surfaces play crucial role in analyzing the black hole uniqueness \cite{Yazadjiev:2015hda,Rogatko:2016mho,Cederbaum:2015fra,Cederbaum:2014gva} and area bounds \cite{Shiromizu:2017ego,Feng:2019zzn,Yang:2019zcn}.  

In further investigations it was found that in spacetimes with rotations such surfaces do not exist, but can be generalized to {\it partilaly umbilic surfaces} filling the volumic {\it photon regions}. A novel mathematical treatment was created \cite{Kobialko:2020vqf,Kobialko:2021uwy} presenting these surfaces as satisfying umbilicity conditions for part of the tangent vectors specified by a certainly defined impact parameter. With varied impact parameter these surfaces fill the volumic regions where bound photon orbits exist such as spherical orbits in the Kerr metric (see \cite{Teo:2020sey} and references therein).  
In turn, this foliation can be used to  construct Killing tensors of the second rank, which are reducible in slices but non-reducible in the complete manifold \cite{Kobialko:2021aqg}. The integrability conditions for the foliation, generating (conformal) Killing tensors, guarantee that slices of the foliation are photon surfaces. This construction generalizes in a natural way to conformal Killing tensors \cite{Kobialko:2022ozq} and demonstrates a deep connection between the integrability of geodesic equations \cite{Cariglia:2014ysa,Frolov:2017kze} and characteristic photon surfaces. 

This framework was further generalized to {\it massive particle surfaces} which have similar properties for timelike geodesics corresponding to massive particles scattered by black holes or other ultracompact objects \cite{Kobialko:2022uzj,Bogush:2023ojz}. Although flows of massive particles are not directly observed from far away (except probably for neutrino, whose detection is still a big challenge) these surfaces can be observed indirectly by their proper radiation which can can be visible in some cases. But more important application of massive particle surfaces lies in relation to photons propagating in plasma which may surround black holes. In inhomogeneous plasma, in addition to gravitational deflection of light, electromagnetic refraction is also present \cite{Bisnovatyi-Kogan:2010flt,Wagner:2020ihx,Fathi:2021mjc,Kumar:2023wfp} which can be incorporated into a combined lensing theory.
Photons in plasma have an effective mass related to Langmiure plasma frequency depending on electron concentration and thus varying in space. Strong lensing, both refractive and gravitational, of photons near black holes surrounded by plasma was intensively studied recently \cite{Bisnovatyi-Kogan:2010flt,Tsupko:2013cqa,Er:2013efa,Liu:2015zou,Atamurotov:2015nra,Abdujabbarov:2016efm,Huang:2018rfn}. Separability of the corresponding equations of motion in Kerr metric for certain plasma configurations was discussed in \cite{Perlick:2017fio,Bezdekova:2022gib}. 
Thus, propagation of light in plasmic media gives rise to concept of a particle with variable mass, whose motion can be described by Polyakov action with coordinate-dependent mass term. For such an action one can further construct geometrical picture of massive characteristic surfaces similarly to the case of particles with constant mass. Formation of shadows of black holes surrounded by plasma was recently discussed in a number of papers \cite{Perlick:2015vta,Bisnovatyi-Kogan:2017kii,Cunha:2018acu}.

The equation of motion for massive particles lacks the conformal invariance of null orbits leading to some complications. For instance, in stationary axisymmetric spacetimes, photon trajectories are determined only by the ratio of azimuthal angular momentum projection to energy $L/E$, while trajectories of massive particles with mass $m$ are defined by two ratios: $E/m$ and $L/m$. Taking a photo of the optic shadow, each pixel corresponds to its own value of $L/E$. But if the photo captures an image of the massive shadow, each pixel corresponds to one-dimensional family in the parametric space $(E/m, L/m)$ and there is no   sharply  defined shadow boundary. One way to solve this issue is to fix an integral of motion, e.g., the specific energy $E/m$, so the photo captures the image of the shadow of particles with a fixed specific energy $E/m$. This leads to a one-dimensional family of shadows for all possible values of $E/m$. Note that the observed particle energy depends on the observer's four-velocity. Particularly, in stationary axisymmetric spacetime, if two observers move along two different Killing vectors, the energy $E_1$ observed by one of them is equal to the linear combination of the energy $E_2$ and angular momentum projection $L_{2}$ observed by the second observer.

A family of spherical photon orbits with the same ratio $L/E$ forms a web aligning into a \textit{photon surface} (occasionally also called a \textit{fundamental photon surface}, if each value of $L/E$ defines a distinct surface like in the Kerr spacetime). The set of all such compact photon surfaces forms a \textit{photon region}. The photon region does not depend on the observer's four-velocity. However, distinctive feature appears for massive particles when we are attempting to introduce a \textit{massive particle region} similarly to the photon region. Massive particle region does depend on the observer, because it contains a set of massive particle surfaces for a fixed specific energy $E/m$ defined individually for each observer. All these observations serve as the foundation for the development of the geometric theory of massive shadows and regions presented in this paper.

We anticipate the application of our results not only in astrophysical observations and theoretical gravity constructions but also in experimental physics, particularly in analog models of gravity and condensed matter physics. Analog models of gravity involve the study of laboratory systems described by equations similar to those in General Relativity \cite{barcelo2011analogue}. These systems can encompass sound and fluid waves, oscillating bubbles in sonoluminescence, photons in media with variable refractive indices, Bose-Einstein condensates, helium super-fluids, slow photons in fluids, and more. Analog models are employed to analyze phenomena such as Hawking radiation \cite{Drori:2018ivu,Carusotto:2008ep}, particle creation \cite{Schutzhold:2007mx,Karmakar:2023yce}, quasi-normal modes \cite{Torres:2020tzs}, quantum particles \cite{Schmidt:2023mjg}, and quantum fields \cite{Viermann:2022wgw} in curved spacetime. Our results can facilitate the analysis of analog models of gravity, pushing forward the understanding of the dynamics of waves through tools like the WKB approximation. In the realm of condensed matter physics, crystal defects can be described using differential geometry \cite{kroner1990differential,clayton2005geometric} giving the connection to the gravity. Moreover, in systems where the potential energy is added to the (quasi-)particle mass, they both together can be combined into one effective variable mass. Also, two dimensional systems can incorporate non-trivial geometry, contributing to the (quasi-)particle dynamics \cite{da1981quantum,grieser2009perturbation,du2022linearized}. We see our tool in the application to studies how particles like phonons, photons, or plasmons move near crystal defects or in curved low-dimensional systems.

The article is organized as follows. In Sec. \ref{sec:setup}, we determine the observer's tetrad and the main observable quantities that affect the structure of the gravitational shadow. We also recall the main features of the geometric description of the massive particle surfaces/region and slice-reducible Killing tensors. In Sec. \ref{sec:MPS}, we derive a general, explicitly coordinate-independent expression for the shadow boundaries, including the case of particles with variable mass and photons in non-magnetized pressureless plasma. In particular, we demonstrate that essentially the properties of the shadow are concentrated in so-called shadow matrix. We also consider asymptotic limit of all expressions to obtain compact formulas for the most relevant observer distant from the black hole. In Sec. \ref{sec:examples}, we provide all results in coordinates of separation variables such as Boyer-Lindquist coordinates and consider general metrics in the Benenti-Francaviglia form and others. Finally, we illustrate the obtained expressions for the shadow boundary using Kerr-NUT and EMD black hole models and compare them with well-known results to confirm correctness and universality.

\section{Setup}
\label{sec:setup}

We assume that the four-dimensional spacetime is  stationary and axisymmetric with two Killing vectors ${\kappa_a}^\alpha$, where $a=0,1$ is enumerating them index. One can define the Gram matrix $\G_{ab} = {\kappa_a}^\alpha {\kappa_b}_\alpha$ and its inverse $\G^{ab} = (\G_{ab})^{-1}$, imposing invertibility. We will use $\G_{ab}$ and $\G^{ab}$ to lower and raise indices associated with the Killing subspace. The set of vectors ${\kappa_a}^\alpha$ can be understood as a matrix projecting any vector to the subspace spanned by Killing vectors. The signature of $\G_{ab}$ assumed to be $(-+)$  to span one timelike and one spacelike directions. For example, it is common to use ${\kappa_0}^\alpha \partial_\alpha = \partial_t$ and ${\kappa_1}^\alpha \partial_\alpha = \partial_\phi$. The Levi-Civita connection $\nabla_\alpha$ acts on subscripts and superscripts $a$, $b$ as a partial derivative $\partial_\alpha$ \cite{Kobialko:2022ozq}, since they represent contractions with a Killing vector labeled by $a$ or $b$, but not tensor components. 

\subsection{Observables}

\begin{figure}
    \centering
    \includegraphics[width=0.75\linewidth]{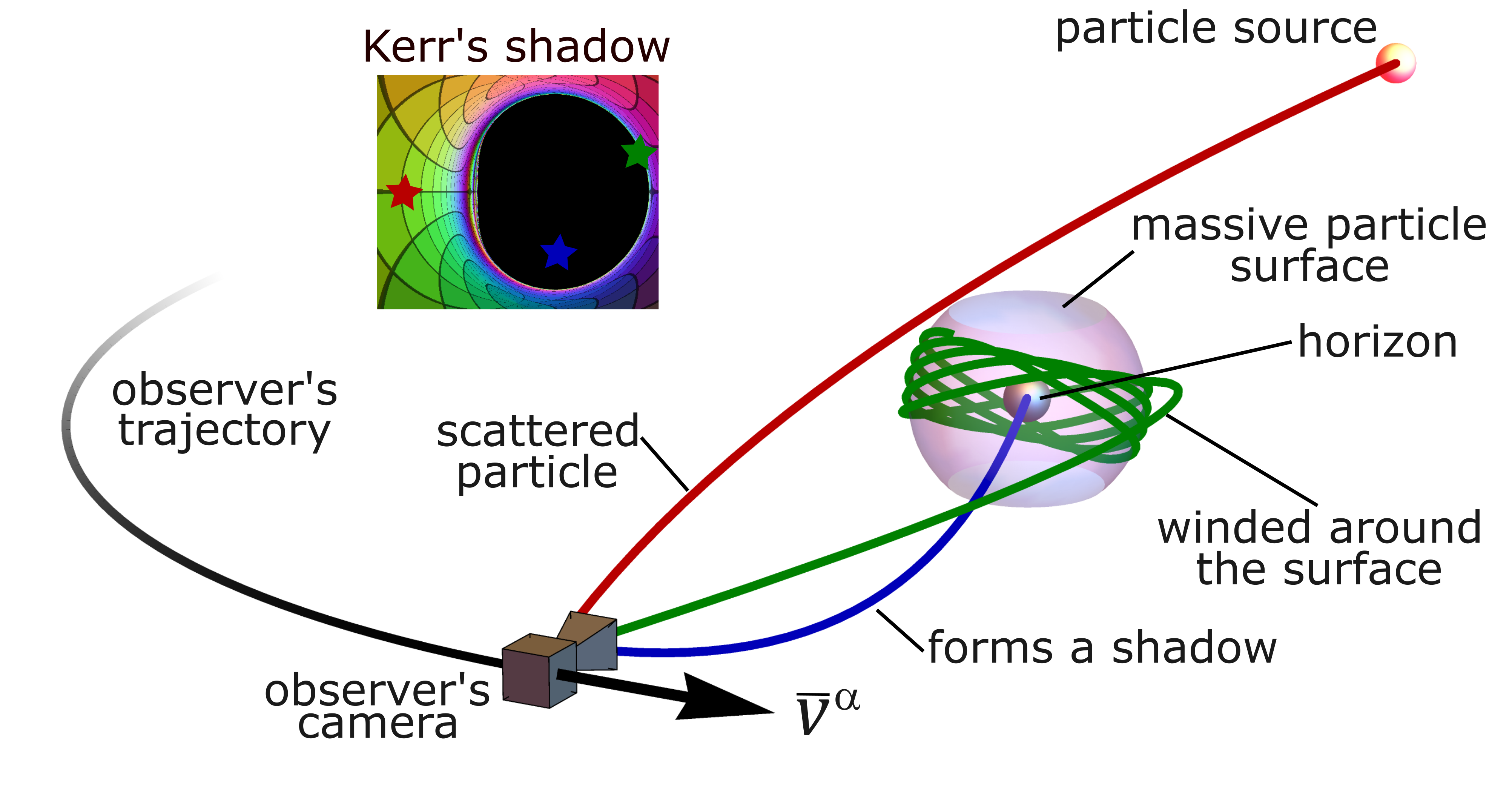}
    \caption{Schematic representation of an observer orbiting a black hole. The blue and red curves depict the geodesics associated with the shadow and scattered particles, respectively. The green curve represents the geodesic corresponding to the boundary of the shadow, winding along the  massive particle surface. The inset schematically shows the Kerr shadow points corresponding to each of these curves.}
    \label{fig:model}
\end{figure}

We define a ``stationary'' \cite{Pugliese:2018hju} observer $O$ with four-velocity $\bar{v}^\alpha = \bar{v}^a \bar{\kappa}_a{}^\alpha$ (Fig. \ref{fig:model}). We will use bars for those quantities, that are calculated at observer's position. Accordingly to the rule of lowering Killing indices, we have
\begin{equation}\label{eq:condition_3_q}
\bar{v}_a = \bar{\G}_{ab}\bar{v}^b = \bar{\kappa}_a{}_\alpha \bar{v}^\alpha.
\end{equation}
If the observer follows the geodesic motion, $ \bar{v}_a$ are the associated conserved quantities.

Now, consider a set of all possible geodesics $\gamma(s)$ captured by the observer following the worldline parametrized with an affine parameter $s$ such that $s=0$ corresponds to observer's point $O$. The set of tangent vectors to such geodesics can be parametrized as follows \cite{grenzebach2016shadow,Briozzo:2022mgg,Bogush:2022hop,Perlick:2023znh}   
\begin{align} \label{eq:gamma}
    \dot{\gamma}^\alpha(0) = 
          \mathfrak{N}\bar{e}_0{}^\alpha
        + \sqrt{\mathfrak{N}^2-m^2} \left(\sin\Phi\sin\Theta \, \bar{e}_1{}^\alpha
        + \cos\Phi\sin\Theta \, \bar{e}_2{}^\alpha
        +\cos\Theta \, \bar{e}_3{}^\alpha\right),
\end{align}
where $\dot{\gamma}^\alpha(s)$ denotes the derivative of $\gamma(s)$ with respect to the affine parameter $s$, $\mathfrak{N}$ is some function to be determined, $\bar{e}_i{}^\alpha$ is an orthonormal tetrad. The term $\sqrt{\mathfrak{N}^2-m^2}$ is introduced in order to ensure the correct norm ${\dot{\gamma}^\alpha\dot{\gamma}_\alpha=-m^2}$. The angles $\Phi\in[0,2\pi]$ and $\Theta\in[0,\pi]$ encode the azimuth and altitude of the observer's celestial sphere, respectively. On the celestial sphere, one can define the zenith ($\Theta=0$), the astronomical horizon ($\Theta=\pi/2$), and the nadir ($\Theta=\pi$). Instead, stereographic projection coordinates can be introduced by the following transformation  \cite{grenzebach2016shadow,Briozzo:2022mgg}  
\begin{equation}
X=-2\tan(\Theta/2)\sin\Phi, \quad Y=-2\tan(\Theta/2)\cos\Phi.
\end{equation}
In what follows, we choose the first two vectors of the orthonormal tetrad to lie in the Killing vector space as
\begin{align} \label{eq:e0_e1}
    \bar{e}_0{}^\alpha = \bar{\kappa}_a{}^\alpha \bar{v}^a/\bar{v}, \qquad
    \bar{e}_1{}^\alpha = \bar{\kappa}_a{}^\alpha\bar{\tau}^a/\bar{v}, 
\end{align}
where
\begin{align} \label{eq:e0_e1_d}
    \bar{\tau}^a \equiv \bar{\G}^{ab}\bar{E}_{bc}\bar{v}^c, \quad
    \bar{v} \equiv \sqrt{-\bar{\G}_{ab}\bar{v}^a \bar{v}^b}, \quad 
    \bar{E}_{ab}\equiv(-\det \bar{\G}_{ab})^{1/2} \epsilon_{ab}, \quad \quad \epsilon_{01}=-\epsilon^{01}=1, 
\end{align}
while the remaining two vectors $\bar{e}_2{}^\alpha$, $\bar{e}_3{}^\alpha$ will be specified soon. Indeed, the orthonormality  properties $\bar{e}_0{}^\alpha \bar{e}_1{}_\alpha=0$ and $-\bar{e}_0{}^\alpha \bar{e}_0{}_\alpha=\bar{e}_1{}^\alpha \bar{e}_1{}_\alpha=1$ follows from Eqs. (\ref{eq:e0_e1}) and (\ref{eq:e0_e1_d}). 

As it was discussed above, due to the lack of conformal invariance, we have to fix some parameters of the observable particle flux. Namely, we define the local observable energy $E$ of the particles as
\begin{align} 
    \bar{E}\equiv - \bar{v}_\alpha  \dot{\gamma}^\alpha(0).
\end{align}
This value $\bar{E}$ is the energy seen by an observer with the four-velocity $\bar{v}^\alpha$, and it is positive, $\bar{E}>0$, since the tangential velocities of the particle and the observer are future-directed. One can define the energy globally as ${E \equiv -  \bar{v}^a {\kappa_a}_\alpha \dot{\gamma}^\alpha(s)}$,
which is conserved along the geodesic motion, and  coincides with the  observable energy $\left.E\right|_{s=0} = \bar{E}$ at the observer's position. Of course, if we considered the case of $\bar{v}^\alpha$ not along the Killing vectors, we would not be able to introduce such a globally defined conserved energy. Fixing the observable energy allows us to construct images of shadows for massive particle fluxes in a reasonable way from experimental perspectives. The entire spectrum of energies will create superposition of these individual images. Once we have fixed the energy, we can determine the  function $\mathfrak{N}$:
\begin{align} 
  \bar{E} = - \bar{v}_\alpha \dot{\gamma}^\alpha(0)
  = - \mathfrak{N}\bar{v}^\alpha  \bar{v}_\alpha / \bar{v}
  = \bar{v} \mathfrak{N}
  \quad \Rightarrow \quad
  \mathfrak{N}=\bar{E}/\bar{v}.
\end{align}

The entire set of constants of motion associated with the Killing vectors ${\kappa_a}^\alpha$ that are conserved along the geodesics can be obtained in the form \cite{Bogush:2023ojz}
\begin{align} \label{eq:rho}
    q_a\equiv \kappa_a{}_\alpha\dot{\gamma}^\alpha(s) = \frac{m}{m_E }\bar{v}_a/\bar{v}+ \frac{m}{m_E }\sin\Phi\sin\Theta \sqrt{1-m^2_E}\bar{\tau}_a/\bar{v}, \quad
    m_E\equiv \bar{v}m/\bar{E},
\end{align}
where  we used Eq. (\ref{eq:gamma}) with Eq. (\ref{eq:e0_e1}), and we remind that $\bar{\tau}_a = \bar{\G}_{ab}\bar{\tau}^b$. Thus, for geodesics with a given observable energy $\bar{E}$ for the given observer $\bar{v}^a$, we have obtained a family of conserved quantities parametrized by $\Phi,\Theta$. The family is one-dimensional since the observable energy $\bar{E}$ is already fixed, imposing a linear condition $q_a \bar{v}^a = -\bar{E}$. Since, the particle's four velocity is a timelike future-directed vector, the following natural inequality on the observable energy arises
\begin{align} \label{eq:condition_mv}
0\leq m_E \leq 1 \quad \Leftrightarrow \quad \bar{E}\geq m\bar{v} \geq 0.
\end{align}
The mass $m$ represents the rest energy of a particle in a static asymptotic observer's reference frame, while the quantity $mv$ represents the rest energy observed by a non-static or non-asymptotic observer. Thus, condition (\ref{eq:condition_mv}) means that the energy of the particle is greater than the rest energy with respect to the observer's reference frame. The parameter $m_E$, contained in the range $0\leq m_E \leq 1$, is convenient to describe the entire energy spectrum.

In what follows, two special types of observers will be important (defined up to some norm)
\begin{itemize}
    \item static $\bar{v}_{\text{st}}^a \sim (1,0)$
    \item zero angular momentum observer (ZAMO) $\bar{v}_{\text{ZAMO}}^a \sim (\bar{\G}_{11},-\bar{\G}_{01})$
\end{itemize}
For asymptotically flat spacetimes, ZAMO observer at the asymptotically distant sphere  approaches the static observer ${\bar{v}_{\text{ZAMO}}^a \to \bar{v}_{\text{st}}^a}$ since $\G_{01} = O(1/r)$. However, as we will show, terms of the order $O(1/r)$ can be important for the shadow.

\subsection{Massive particle surfaces} 

Similarly to the case of photons, Refs. \cite{grenzebach2016shadow,Briozzo:2022mgg,Perlick:2023znh,Bogush:2022hop,Grenzebach:2014fha,Frost:2023enn,Grenzebach:2015oea,Virbhadra:1999nm}, the boundary of the massive particle shadow will be formed by particles which are asymptotically tangent to the massive particle surfaces with a compact spatial section. The formal definition of massive particle surfaces is given in Refs. \cite{Kobialko:2022uzj,Bogush:2023ojz} (also, see Ref. \cite{Song:2022fdg}). In simple words, a hypersurface $S$ is the massive particle surface if any worldline with a given set of conserved quantities $q_a$ that touches $S$ (at least at one point) belongs to $S$ entirely. As shown in Ref. \cite{Bogush:2023ojz}, the second fundamental form (extrinsic curvature) of massive particle surfaces for neutral particles with a set of conserved quantities $q_a$ must satisfy the equation
\begin{equation}\label{eq:condition_chi}
\chi_{\alpha\beta} =
    \chi_\tau \left(
          h_{\alpha\beta}
        + \mathcal{H}^{ab} \kappa_{a\alpha} \kappa_{b\beta}
    \right),  
\end{equation}
where $h_{\alpha\beta}$ is the induced metric, $\chi_\tau$ is an arbitrary function and $\mathcal{H}^{ab}$ is an arbitrary matrix restricted only by the following constraint
\begin{align} \label{eq:condition_Hq}
    \mathcal{H}^{ab} q_{a} q_{b}= m^2,
\end{align}
and the following inequality for geodesic motion
\begin{align} \label{eq:condition_Gq}
    \G^{ab}q_a q_b + m^2 \leq 0,
\end{align}
and all points of the massive particle surface $S$ must satisfy this condition. As $q_a$ represents a geodesic path passing through the observer's point, it inherently satisfies the inequalities at the observer's point. The maximal connected region containing the observation point is the \textit{observable region} \cite{Kobialko:2020vqf} (see Ref. \cite{Huang:2018rfn} for the case of photons in plasma). The gravitational shadow can be formed only for those particles that are in the observable region containing the black hole's event horizon or other trapping surface. Otherwise, such geodesics will create the boundary of relativistic images \cite{Virbhadra:1999nm}. 

Of our greatest interest will be the axi-stationary massive particle surfaces which are touched by all the Killing vector fields $\kappa_{a}{}^\alpha$, i.e., all Killing vectors $\kappa_{a}{}^\alpha$ are tangent to the massive particle surface $S$. In this case, according to Ref. \cite{Kobialko:2022ozq} the matrix $\mathcal{H}^{ab}$ is expressed in terms of the Gram matrix $\G_{ab}$ 
\begin{align}\label{eq:HasG}
    \mathcal{H}^{ab} =- \frac{1}{2\chi_\tau} n^\beta  \nabla_\beta\G^{ab} - \G^{ab}.
\end{align}
We will also consider only the massive particle surfaces with a compact spatial section, just like in the case of fundamental photon surfaces \cite{Kobialko:2020vqf}. 

The family of conserved quantities $q_a$ defined in the previous section corresponds to a family of massive particle surfaces. We will call this family  for all possible $\Phi$ and $\Theta$ the \textit{massive particle region}. In the case of photons, an individual photon is emitted by some source, asymptotically approaches the corresponding photon surface and moves away from it in the direction of the observer \cite{grenzebach2016shadow,Grenzebach:2014fha}. Similarly to the case of photons, scattering of massive particles will occur on individual surfaces in the massive particle region. Thus, this definition is key to the problem of constructing the gravitational shadow and relativistic images.

\subsection{Killing tensors}   

In the general case, the problem of finding the massive particle region is very nontrivial. However, it is significantly simplified in the case of integrable systems \cite{Konoplya:2021slg,Frolov:2017kze}. Assume that there is an exact slice-reducible Killing tensor in 
 the spacetime, defined in Refs. \cite{Kobialko:2021aqg,Kobialko:2022ozq}. Recall that Killing tensor is called \textit{slice-reducible} if there exists at least one foliation of the spacetime such that tangent projections in all slices are reducible. Although, slice-reducible Killing tensors do not represent the most general Killing tensors, they are very common among physical spacetimes. We give here a general theorem about their form and necessary and sufficient conditions for their existence: 

\begin{theorem} \label{KBG_non}
Let the manifold $M$ of dimension $m$ be foliated by slices $S$ and contains a collection of $n\leq m-2$ Killing vector fields $\kappa_a{}^\alpha$ tangent to the foliation slices $S$. Let the Gram matrix $\G_{ab}$ be non-degenerate, and the basis of foliation slices be completed by the set of $m-n-1$ vectors $\sigma_i^\alpha$ orthogonal to Killing vectors (i.e., $\kappa_{a}{}^\alpha {\sigma_i}_\alpha = 0$). The unit vector field $n^\alpha$ is normal to slices $S$. If the second fundamental form of slices possesses the following form
\begin{equation} \label{eq:block_condition}
\SFS_{\alpha\beta}\kappa_{a}{}^\alpha \sigma_i^\beta=0,\qquad
\SFS_{\alpha\beta}\sigma_i^\alpha\sigma_j^\beta=\chi_\tau h_{\alpha\beta}\sigma_i^\alpha\sigma_j^\beta,
\end{equation}
and the following integrability conditions are met
\begin{subequations}\label{eq:integrability_conditions}
\begin{align}
    \label{eq:integrability_1}
   & \LCS_\gamma \left( \varphi \chi_\tau - \varphi n^\alpha \LCM_{\alpha} \ln \varphi \right)
    = 0,\\\label{eq:integrability_2}
   & \LCS_\gamma\left(
    \frac{1}{2\chi_\tau}  n^\alpha \nabla_\alpha \G^{ab}+  \G^{ab}
    \right) =0,\\\label{eq:integrability_3}
   & \LCS_\gamma (\chi_\tau\varphi^3) = 0,
\end{align}
\end{subequations}
then, there is a slice-reducible Killing tensor in the manifold $M$, which can be constructed as follows:

{\bf Step one}: Obtain $\Psi = \ln \varphi^2 + \tilde\Psi$ from the equations
\begin{align} \label{eq:psi}
     n^\alpha \LCM_{\alpha}\tilde{\Psi} =
    2 \left(\chi_\tau - n^\alpha \LCM_{\alpha}\ln \varphi\right), \quad \LCS_\tau \tilde \Psi = 0.
\end{align}

{\bf Step two}: Obtain $\alpha$ and $\gamma^{ab}$ from the equations
\begin{subequations}
    \begin{equation} \label{eq:alpha}
       n^\alpha \LCM_{\alpha} \alpha = - 2  \chi  e^{\Psi}, \quad
       \LCS_\tau \alpha = 0.
    \end{equation}
    \begin{equation}\label{eq:gamma_condition}
        \gamma^{ab} =  e^{\Psi} \G^{ab} + \nu^{ab}, \quad
        n^\alpha\LCM_{\alpha} \nu^{ab} = 0, \quad \LCS_\tau \gamma^{ab} = 0.
    \end{equation} 
\end{subequations}

{\bf Step three}: Using the functions found in the previous steps, construct the corresponding KT:
\begin{align}\label{eq:KT}
   & K_{\alpha\beta} =
      \alpha g_{\alpha\beta}
    + \gamma^{ab} \kappa_{a\alpha} \kappa_{b\beta}
    + e^{\Psi} n_{\alpha} n_{\beta}.  
\end{align} 
\end{theorem}
Here, $\LCS_\alpha$ denotes the covariant derivative tangent to the slices $S$, and $\varphi$ is a lapse function obeying the equation $n^\lambda\nabla_\lambda n_\beta = -\LCS_\beta \ln \varphi$, more details can be found in Sec. II from Ref. \cite{Kobialko:2022ozq}.

In Ref. \cite{Bogush:2023ojz}, the slices generating the Killing tensors are shown to be massive particle surface. Indeed, the second integrability condition (\ref{eq:integrability_2}) can be rewritten in the form $\LCS_{\alpha} \mathcal{H}^{ab}=0$, i.e., the matrix $\mathcal{H}^{ab}$ is constant in each slice (Refs. \cite{Kobialko:2021aqg,Kobialko:2022ozq}). Thus, if equation (\ref{eq:condition_Hq}) has a real solution $q_a$ at some point of the slice, then the part of this slice that satisfies inequality (\ref{eq:condition_Gq}) is automatically a massive particles surface. Even more, this slice is shown to allow for a continuous family of solutions $q_a$ instead of an isolated solution (this case is called a glued surface \cite{Bogush:2023ojz}). If the inequality is not satisfied, e.g., the solution is complex, then the slice is not a massive particle surface.

\begin{figure}
    \centering
    \includegraphics[width=0.35\linewidth]{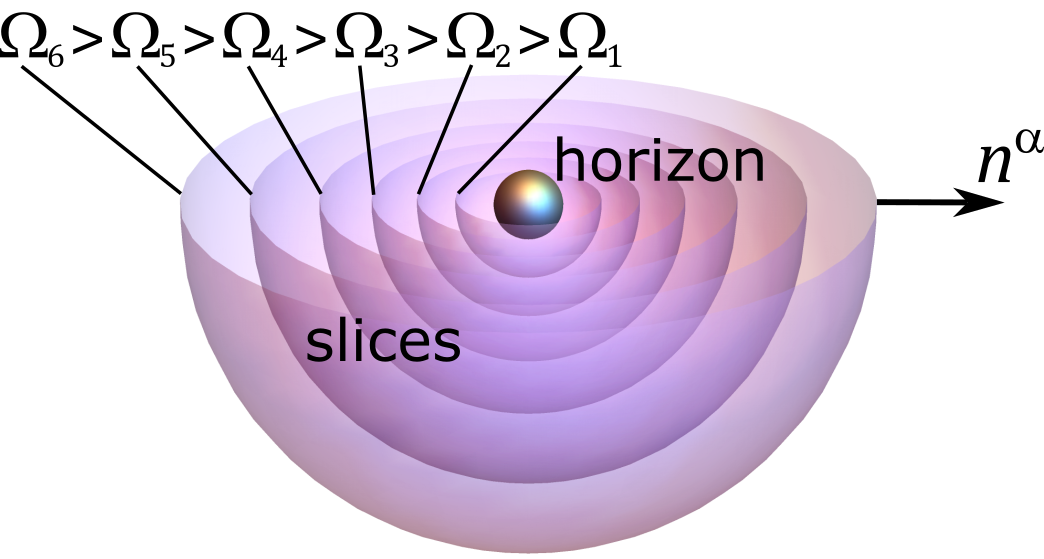}
    \caption{Schematic representation of a foliation by slices of the spacetime. Slices fill the entire spacetime.}
    \label{fig:foliation}
\end{figure}

In order to make massive particle surfaces a meaningful tool for analyzing shadows, we assume the compactness of their spatial section. Otherwise, if the massive particle surface were not compact, the geodesic would have an infinite volume of space to travel. To our knowledge, all physically meaningful four-dimensional solutions that possess a Killing tensor have two foliations with slices satisfying the integrability conditions from the theorem -- the slices with cone-like and sphere-like spatial sections. In this paper we use the sphere-like slices because the spatial section of the corresponding massive particle surfaces are guaranteed to be compact. To be specific, we choose the outer normal $n^\alpha$ pointing to the infinity. As a result the foliation slices can be seen as flowing out of the compact source, filling the entire space (Fig. \ref{fig:foliation}).

Now we can define the {\it massive particle region}. Unlike the previous subsection, it will be more convenient for us to parameterize the family not by  $\Phi$ and $\Theta$, but by the foliation parameter $\Omega$ that generates the Killing tensor (e.g., for many solutions, such a foliation parameter is just the coordinate $r$ in Boyer–Lindquist coordinates \cite{Grenzebach:2014fha,grenzebach2016shadow}). This is motivated by the fact that in most cases the product $\sin\Phi\sin\Theta$ is expressed through some high-order polynomials of the foliation parameter, which cannot be resolved in the opposite direction in radicals \cite{grenzebach2016shadow}.

Since the slices are massive particle surfaces they obviously satisfy Eq. (\ref{eq:condition_Hq}) for some $q_a$. Keeping in mind that using the equations for the slice-reducible tensor, we can get an alternative representation for $\mathcal{H}^{ab}$ \cite{Kobialko:2022ozq}
\begin{align}\label{eq:HtoG}
    \mathcal{H}^{ab}
    =
    -\frac{1}{2 \chi_\tau }n^\alpha \nabla_\alpha \G^{ab}
    -\G^{ab}
    =-\frac{ n^\alpha \nabla_\alpha (e^\Psi \G^{ab})}{n^\beta  \nabla_\beta e^\Psi }
    ,
\end{align}  
which allows us to rewrite equation (\ref{eq:condition_Hq}) as
\begin{align}\label{eq:HtoGm}
 (e^\Psi \G^{ab})' q_a q_b + (e^\Psi m^2)'=0,
\end{align} 
where the prime $'$ means the derivative with respect to the foliation parameter $\varphi n^\alpha \LCM_\alpha$. As in the case of photons, this equation can be represented in the homogeneous form with respect to the \textit{impact parameter} vector $\rho_a$ \cite{Kobialko:2021aqg,Kobialko:2022ozq}
\begin{align} \label{eq:xi_sol}
\left(\mathcal{S}^{ab}\right)'\rho_a \rho_b=0,
\end{align}  
where we have defined an object that will be called \textit{shadow matrix}:
\begin{align} \label{eq:shadow_matrix}
\mathcal{S}^{ab}\equiv e^\Psi \left(\G^{ab} + m^2_E\cdot\bar{v}^a\bar{v}^b/\bar{v}^2 \right), 
\end{align}  
and the original conserved quantities $q_a$ in (\ref{eq:HtoGm}) are expressed through 
\begin{align} \label{eq:q_sol}
q_a=-\frac{m\bar{v}}{m_E}\cdot\rho_a/(\bar{v}^b\rho_b ).
\end{align}
Relation (\ref{eq:q_sol}) is chosen in such a way that the identity $\bar{v}^a q_a = - \bar{E}$ holds automatically. It is seen from Eq. (\ref{eq:q_sol}), that the impact parameter vector $\rho$ has an arbitrary norm. By choosing $\rho_0=-1$, its second component becomes a conventional impact parameter of a geodesic $\rho_1=-q_1/q_0$. According to Eq. (\ref{eq:gamma_condition}), the derivative along directions tangent to the slices is also zero
\begin{equation}\label{eq:m_integrability_1}
    \LCS_\gamma\left[\left(\mathcal{S}^{ab}\right)'\right]=0.
\end{equation}
Furthermore, comparing Eqs. (\ref{eq:q_sol}) and (\ref{eq:rho}) and contracting with vector $\bar{\tau}^a$, one will find the solution for sines:
\begin{align}  \label{eq:sinsin}
\sin\Phi\sin\Theta=-\frac{1}{\sqrt{1-m^2_E}}\cdot(\bar{\tau}^a\rho_a)/(\rho_b \bar{v}^b).
\end{align}
Obviously, Eq. (\ref{eq:xi_sol}) does not always have solutions which also satisfies Eq. (\ref{eq:condition_Gq}). Thus some slices of the foliation does not contain massive particle surfaces with a given value of the observed energy $\bar{E}$. The following existence conditions can be identified. First, the matrix $\left(\mathcal{S}^{ab}\right)'$ must be indefinite, that is, its determinant must be non-positive $\det\{\left(\mathcal{S}^{ab}\right)'\}\leq0$. The second condition follows from the natural inequality $|\sin\Phi\sin\Theta|\leq1$
\begin{align} \label{eq:rho_c}
\left(\bar{\tau}^a\rho_a\right)^2\leq\left(1-m^2_E\right)\left( \bar{v}^b\rho_b\right)^2.
\end{align}
The last condition follows from (\ref{eq:condition_Gq}), which can be rewritten as (assuming $e^\Psi>0$)
\begin{align} \label{eq:MPR}
\mathcal{S}^{ab}\rho_a \rho_b\leq0.
\end{align}  

Summarizing, when a slice-reducible Killing tensor exists, we have the following simple coordinate independent recipe for constructing a massive particle region. For each slice with compact spatial section on which the Killing tensor is reducible, find the corresponding shadow matrix $\mathcal{S}^{ab}$ and $\rho_a$ if the latter exists. Then, find the part of the slice that satisfies the condition $\mathcal{S}^{ab}\rho_a \rho_b\leq0$. The union of all such submanifolds of all slices constitutes the massive particle region. This time the massive particle region is parameterized by the foliation parameter $\Omega$. In the case of photons, it will exactly coincide with the well-known photon region, Refs. \cite{Perlick:2021aok,Grenzebach:2014fha,grenzebach2016shadow,Grenzebach:2015oea}. 

The advantage over the classical approach of solving geodesic equations for construction of massive particle region is that we no longer need to go through the procedure of solving geodesic equations by explicitly selecting a suitable coordinate system at all. For example, in the Kerr's case we do not need to work in the Boyer–Lindquist coordinate system (or any other coordinate system that provides separability) to define the massive particle region. The only job we have to do is to find slices on which the Killing tensor becomes reducible. In the general case, such slices may differ from the standard surfaces $r=\text{const}$. This can be particularly advantageous in spacetimes where the coordinates that provide separation of variables are unknown, or when the non-separating coordinates yield more concise expressions. The only necessary and sufficient conditions that allow applying the results of this paper are the integrability conditions (\ref{eq:integrability_conditions}) along with (\ref{eq:block_condition}).

\section{Massive particle shadows}
\label{sec:MPS}

\subsection{Basic definitions}

Having defined the massive particle region, we can take the final step towards obtaining a formula for the boundary of the gravitational shadow, or at least the boundary of relativistic images. Just established relationship between geodesics that go through the observer $O$ with a fixed $\rho_a$ and the corresponding massive particle surface $S$ can be interpreted by considering the four following cases, tracking the geodesics in the inverse direction (i.e., from the observer to the source) \cite{grenzebach2016shadow,Grenzebach:2015oea,Grenzebach:2014fha,Virbhadra:1999nm,Frost:2023enn}:
\begin{enumerate}[\itshape(i)]
    \item If a geodesic goes far away from the surface, it flies-by the massive gravitating object with a small distortion of its path (red curve in Fig. \ref{fig:model}).
    \item If a geodesic is about to touch the surface, it wraps around the surface several times and fly away. This is the effect of the relativistic images formation \cite{Virbhadra:1999nm,Tsupko:2013cqa}.
    \item If a geodesic is infinitely close to touching the surface, it wraps the surface an infinite number of times (green curve in Fig. \ref{fig:model}). It corresponds to the boundary of the shadow.
    \item If a geodesic intersects the surface, it falls inside the surface and get trapped by the massive gravitating object (blue curve in Fig. \ref{fig:model}). As it will be mentioned below, this may be not fair for some objects like naked singularities. In the case of   regular black holes without exotic matter, all the geodesics that get inside the surface are caught by the horizon forming a dark spot -- a gravitational shadow of the black hole.
\end{enumerate}
If the scenario \textit{(iv)} is realized, the boundary of the gravitational shadow coincides with the boundary of relativistic images described in the scenario \textit{(iii)}. This is true if we assume that the geodesic approaching the object closer and closer suffers a stronger gravitational attraction. However, in the case of superextremal solutions, naked singularities and wormholes, the scenario \textit{(iv)} can fail. In particular, geodesics can also turn back inside the massive particle surfaces. In the latter case, light spots may be observed inside the shadow or the shadow may disappear completely, but the boundary of the relativistic images remains unchanged \cite{Bogush:2022hop}. 

Of course, to determine the entire boundary of the shadow, it is not enough to know the vector $\rho_a$. Fortunately, the presence of the Killing tensor allows us to establish another connection between the massive particle surface and the observed parameters of the particle through the conserved Carter's constant \cite{Walker:1970un,Carter:1968ks,Carter:1977pq}
\begin{equation}
\mathcal{C}\equiv K_{\alpha\beta}\dot{\gamma}^\alpha(s)\dot{\gamma}^\beta(s)=-\alpha m^2
    + \gamma^{ab} q_a q_b
    + e^{\Psi} \left(n_{\alpha}\dot{\gamma}^\alpha(s)\right)^2,
\end{equation}
where we have applied Eq. (\ref{eq:KT}) for slice-reducible Killing tensors.
By calculating this expression at the observation point and at the arbitrary point of geodesic tangent to the massive particle surface with a given set of conserved quantities $q_a$ we get
\begin{equation}
-\bar{\alpha} m^2 + \bar{\gamma}^{ab}q_a q_b + (m/m_E)^2(1-m^2_E) e^{\bar{\Psi}} \cos^2\Theta =  -\alpha m^2 + \gamma^{ab} q_a q_b,
\end{equation}
where we have chosen the tetrad vector $\bar{e}_3{}^\alpha=-n^\alpha$ (since we choose the outer normal vectors of the surface, observer's zenith points towards the black hole). Substituting explicitly Eq. (\ref{eq:q_sol}) gives the following equation
\begin{equation} \label{eq:cos_1}
(1-m^2_E) e^{\bar{\Psi}} \cos^2\Theta =  -(\alpha-\bar{\alpha}) m^2_E +\bar{v}^2(\gamma^{ab}-\bar{\gamma}^{ab}) \cdot\rho_a\rho_b/(\bar{v}^c\rho_c)^2.
\end{equation}
According to Ref. \cite{Kobialko:2022ozq}, matrix $\gamma^{ab}$ can be decomposed onto the following parts
\begin{subequations}
    \begin{align}
    \label{eq:gamma_1}
    &
        \gamma^{ab} = e^{\Psi} \G^{ab} + \nu^{ab},
    \\ \label{eq:gamma_2}
    & 
        n^\alpha\LCM_{\alpha} \nu^{ab} = 0,
    \\ \label{eq:gamma_3}
    & 
        \LCS_\tau \gamma^{ab} = 0.
    \end{align} 
\end{subequations}
\begin{figure}
    \centering
    \includegraphics[width=0.7\linewidth]{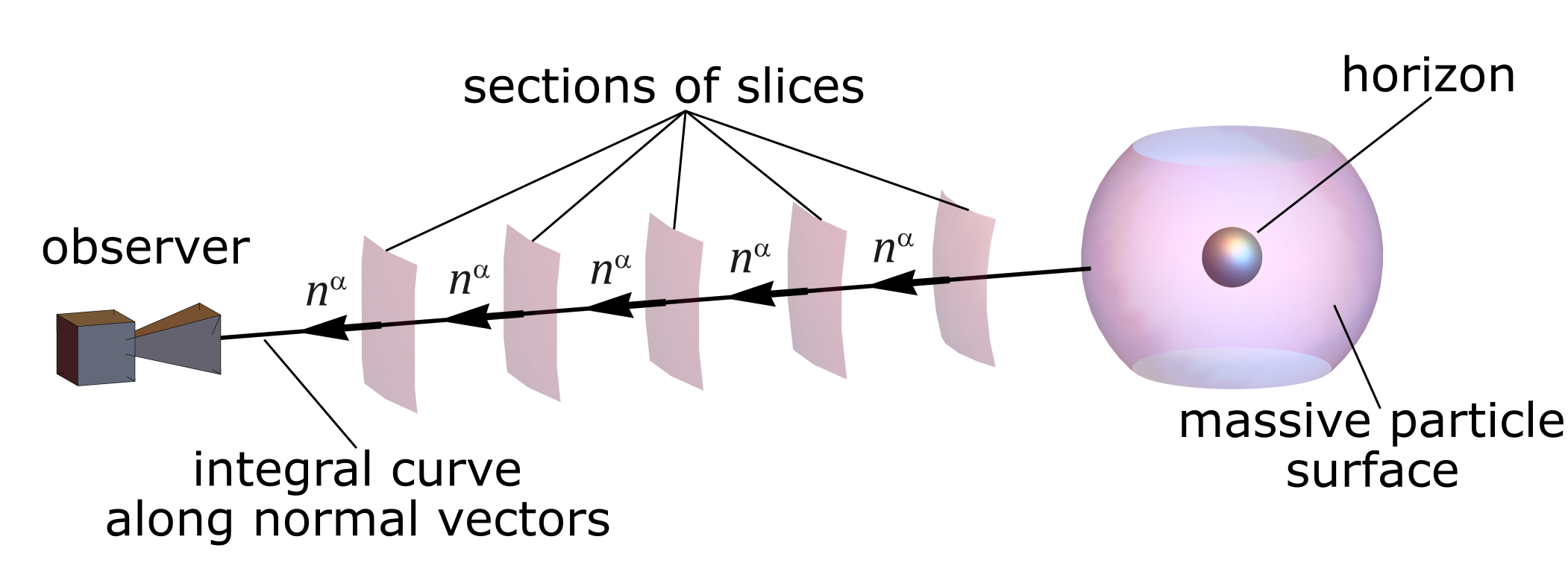}
    \caption{Schematic representation of an integral curve along normal vectors of slices from the observer to the massive particle surface. In the general case, the integral curve may be not straight line.}
    \label{fig:integral_curve}
\end{figure}
The condition (\ref{eq:gamma_3}) states that $\gamma^{ab}$ is constant on the surface. In particular, the expression (\ref{eq:cos_1}) does not actually depend on the point on the massive particle surface. However, according to the condition (\ref{eq:gamma_2}), the term $\nu^{ab}$ does not change along the integral curves of the normal vector field $n^\alpha$. Thus, it is reasonable to evaluate the expression on the surface at the point, which is connected with the observer by the integral curve (Fig. \ref{fig:integral_curve}). Though, it may sound difficult for general foliations, in practice we deal with foliations with slices defined by the constant radius in Boyer-Lindquist coordinates \cite{grenzebach2016shadow}, so two points are connected by the integral curve along $n^\alpha$ if they have the same coordinates at their surfaces. So the following difference can be written as
\begin{align} \label{eq:gamma_diff}
(\gamma^{ab}-\bar{\gamma}^{ab})\rho_a \rho_b= \left(e^{\Psi} \G^{ab}- e^{\bar{\Psi}} \bar{\G}^{ab}\right)\rho_a \rho_b,
\end{align}  
where now the expressions without a bar must be evaluated in the slice point which lies in the orbit of observer positions generated by integral curve of normals $n^\alpha$ (Fig. \ref{fig:integral_curve}).

The sum $\alpha + e^\Psi$ is constant along the integral curves along $n^\alpha$ either. To check this, we act on the sum with the derivative along $n^\alpha$:
\begin{align}
    n^\alpha\LCM_\alpha\left(\alpha+e^{\Psi}\right)
    =
    - 2  \chi_\tau  e^{\Psi}+e^{\Psi}n^\alpha\LCM_\alpha \Psi
    =
    - 2  \chi_\tau  e^{\Psi}+e^{\Psi}\cdot (2  \chi_\tau)
    =0,
\end{align}
where we take into account Eqs. (\ref{eq:alpha}), (\ref{eq:psi}). This sum, calculated at the observer's position, $\bar{\alpha} + e^{\bar{\Psi}}$, can be compared with its value ${\alpha} + e^{{\Psi}}$ at any other arbitrary point along the integral curve along the normal vector $n^\alpha$:
\begin{align} \label{eq:alpha_psi}
\alpha+e^{\Psi}-(\bar{\alpha} +e^{\bar{\Psi}})=0 \quad \Rightarrow \quad \alpha-\bar{\alpha}=e^{\bar{\Psi}}-e^{\Psi}.
\end{align}
Substituting Eqs. (\ref{eq:gamma_diff}) and (\ref{eq:alpha_psi}) into Eq. (\ref{eq:cos_1}), we get an equivalent expression
\begin{align} \label{eq:cos_2}
(1-m^2_E) e^{\bar{\Psi}} \cos^2\Theta &=  -(e^{\bar{\Psi}}-e^{\Psi}) m^2_E +\bar{v}^{2}\left(e^{\Psi} \G^{ab}- e^{\bar{\Psi}} \bar{\G}^{ab}\right) \frac{\rho_a\rho_b}{(\bar{v}^c\rho_c)^2}\nonumber\\&=\bar{v}^{2}\left(\mathcal{S}^{ab}-\overline{\mathcal{S}}{}^{ab}\right)\frac{\rho_a\rho_b}{(\bar{v}^c\rho_c)^2},
\end{align}
where quantities with no bars are calculated at the point lying at the same integral curve as the observer does. This expression allows us to get $\cos\Theta$ in terms of $\bar{\Psi}$, $\mathcal{S}^{ab}$ and $\rho_a$ (taken in observer's point and the point on the massive particle surface connected with the observer by the integral normal curve). Collecting Eqs. (\ref{eq:cos_2}), (\ref{eq:sinsin}) together gives the final expressions for the shadow boundary in coordinates $(\Phi,\Theta)$  
\begin{align} \label{eq:Theta_Phi}
    \cos\Theta =\pm
    \frac{\bar{v}e^{-\bar{\Psi}/2} \sqrt{
          \left(\mathcal{S}^{ab}-\overline{\mathcal{S}}{}^{ab}\right)\rho_a\rho_b}}{\sqrt{1-m^2_E}\cdot(\bar{v}^c\rho_c)},
    \quad
    \sin\Phi \cdot \sin\Theta
    =-\frac{(\bar{\tau}^a\rho_a)}{\sqrt{1-m^2_E}\cdot(\rho_b \bar{v}^b)}.
\end{align}
In terms of the stereographic projection coordinates $(X,Y)$, the shadow boundary reads
\begin{align}
    X=-\frac{2 B}{1\mp A}, \quad
    Y=\pm\frac{2\sqrt{1-A^2-B^2}}{1\mp A}, \quad 
    A \equiv \left|\cos\Theta\right|, \quad
    B \equiv \sin\Phi \cdot \sin\Theta.
\end{align}
The expression from the square root has a simple form
\begin{align}
1-A^2-B^2&=1- \frac{\bar{v}^{2}e^{-\bar{\Psi}} 
          \left(\mathcal{S}^{ab}-\overline{\mathcal{S}}{}^{ab}\right)\rho_a\rho_b+(\bar{\tau}^a\rho_a)^2}{(1-m^2_E)\cdot(\bar{v}^c\rho_c)^2}=-\frac{\bar{v}^{2}e^{-\bar{\Psi}} 
          \mathcal{S}^{ab}\rho_a\rho_b}{(1-m^2_E)\cdot(\bar{v}^c\rho_c)^2}, 
\end{align}
where we have used the identity 
\begin{align} \label{eq:fgr}
\overline{\mathcal{S}}{}^{ab}\rho_a\rho_b=-e^{\bar{\Psi}}(1-m^2_E)\cdot(\bar{v}^c\rho_c)^2/\bar{v}^{2}+e^{\bar{\Psi}}(\bar{\tau}^a\rho_a)^2/\bar{v}^{2}, 
\end{align}
that follows from orthogonality of $\bar{v}^a$ and $\bar{\tau}^c$:
\begin{align}
\overline{\mathcal{G}}{}^{ab}\rho_a\rho_b= -(\bar{v}^a\rho_a)^2/\bar{v}^{2}+(\bar{\tau}^a\rho_a)^2/\bar{v}^{2}.
\end{align}
Thus, the final expression for the shadow boundary is
\begin{subequations}\label{eq:result_1}
 \begin{align} 
X&=\frac{2\bar{\tau}^a\rho_a}{\sqrt{1-m^2_E}\cdot\rho_a \bar{v}^a\mp \bar{v}e^{-\bar{\Psi}/2} \sqrt{
          \left(\mathcal{S}^{ab}-\overline{\mathcal{S}}{}^{ab}\right)\rho_a\rho_b}}, \\
Y&=\pm\frac{2\bar{v}e^{-\bar{\Psi}/2} \sqrt{-\mathcal{S}^{ab}\rho_a\rho_b}}{\sqrt{1-m^2_E}\cdot\rho_a \bar{v}^a\mp \bar{v}e^{-\bar{\Psi}/2} \sqrt{
          \left(\mathcal{S}^{ab}-\overline{\mathcal{S}}{}^{ab}\right)\rho_a\rho_b}}.
\end{align}
\end{subequations}
In Eq. (\ref{eq:result_1}), the shadow boundary is described parametrically by the foliation parameter. The foliation parameter takes all values corresponding to the massive particle region. We will come back to the discussion of this result after we generalize it to the case of variable mass.

\subsection{Particles with variable mass and separability}
 
It is known that photons in non-magnetized pressureless plasma acquire an effective description by the Hamiltonian \cite{Perlick:2015vta,Perlick:2017fio,Briozzo:2022mgg,Perlick:2023znh,Bezdekova:2022gib}:
\begin{align} \label{eq:HM}
H=\frac{1}{2}\left(g^{\alpha\beta}\pi_\alpha \pi_\beta + \omega^2_p\right),
\end{align}
where $\omega_p$ is the plasma electron frequency which equals, up to a scalar factor, the electron density. Plasma frequency $\omega_p$ in the Hamiltonian (\ref{eq:HM}) can be interpreted as effective mass $m=\omega_p$ of the photons. Since plasma generically is inhomogeneous, the effective mass can vary in spacetime, i.e., photons in plasma are described as particles with variable mass. Other geodesic systems with variable mass are presented by particles interacting with the scalar field, dark matter with time-dependent mass \cite{Anderson:1997un}, or an effective description of geodesics in higher dimensions (see App. B in Ref. \cite{Bogush:2020obx} and Ref. \cite{Tang:2022bcm}). Unlike the previously noted literature, instead of the Hamiltonian formalism, we give preference to second-order equations of motion that are closely related to the massive particle surfaces. 

The action for particles with variable mass can be written in a usual Polyakov form \cite{Polyakov:1981rd,Anderson:1997un}:
\begin{align}\label{eq:action_m}
S=\frac{1}{2}\int \{\sigma^{-1} g_{\alpha\beta}\dot{\gamma}^\alpha \dot{\gamma}^\beta -m^2 \sigma\} ds,
\end{align}
where $\sigma$ is a Lagrange multiplier and the mass $m$ is considered to be some prescribed  coordinate dependent scalar function $m(x)$. The Hamiltonian for (\ref{eq:action_m}) reads
\begin{align} 
H=\frac{\sigma}{2}\left(g^{\alpha\beta}\pi_\alpha \pi_\beta + m^2\right), \quad \pi_\alpha = \sigma^{-1} g_{\alpha\beta} \dot{\gamma}^\beta.
\end{align}
Variation of the action with respect to $\sigma$ gives the constraint
\begin{align}
g_{\alpha\beta}\dot{\gamma}^\alpha \dot{\gamma}^\beta=-m^2\sigma^2.
\end{align}
The parameterization fixing $\sigma=1$ leads to the Hamiltonian (\ref{eq:HM}) and the geodesic equations with an effective gradient force at the right-hand side:
\begin{align}\label{eq:geodesic_equation_var_m}
\dot{\gamma}^\alpha\LCM_\alpha \dot{\gamma}^\beta=- m g^{\beta\lambda} \LCM_\lambda m, \quad \dot{\gamma}^\alpha\dot{\gamma}_\alpha=-m^2.
\end{align}
This dynamical system, with similar equations in the  flat space, is considered in various sources, such as Ref. \cite{langmuir1928oscillations} or  lectures on wave propagation in an inhomogeneous plasma.

It is expected that the equilibrium plasma distribution in a stationary axisymmetric gravitational field will inherit these symmetries, so the corresponding mass distributions will also be stationary and axisymmetric. A more subtle question is whether the Killing tensor symmetry of spacetime will ensure the separability of the equations of motion of a particle of variable mass. Particular mass distributions that allow one to generalize the Carter constant were found in the Kerr metric \cite{Perlick:2017fio} and some more general metrics \cite{Bezdekova:2022gib} (see also \cite{Badia:2021kpk,Konoplya:2021slg}) using the Hamilton-Jacobi equation in Boyer-Lindquist coordinates. Here we formulate a coordinate-independent method for searching mass distributions that ensure Killing tensor symmetries of the dynamics of variable mass particles. We do this generalizing technique for slice-reducible exact and conformal Killing tensors in suitably foliated spacetimes.
 
Basic constructions of the previous section and Ref. \cite{Bogush:2023ojz} can be generalized to the case of the variable effective mass in different ways, and the simplest one is to combine the Weyl transformation of the metric tensor with the geodesic reparameterization \cite{Perlick:2017fio}. This will bring us back to the case of constant mass and allow us to apply all previous formulas without changes. Indeed, performing the Weyl transformation $\tilde{g}_{\alpha\beta}=e^{2\psi}g_{\alpha\beta}$ supplemented by the mass and the Lagrange multiplier redefinitions $\tilde{m} = e^{-\psi} m$, and  $\tilde{\sigma}= e^{2\psi}\sigma$ brings the action to the form
\begin{align}
S=\frac{1}{2}\int \left(\sigma^{-1} e^{-2\psi}\tilde{g}_{\alpha\beta}\dot{\gamma}^\alpha \dot{\gamma}^\beta -m^2 \sigma\right) ds =\frac{1}{2}\int \left(\tilde{\sigma}^{-1} \tilde{g}_{\alpha\beta}\dot{\gamma}^\alpha \dot{\gamma}^\beta -\tilde{m}^2 \tilde{\sigma}\right) ds.
\end{align}
We choose the transformation function $\psi$ such that $\tilde{m}$ is some constant (of the same sign as $m$), $\psi = \ln (m / \tilde{m})$. The parameterization choice $\tilde{\sigma}=1$ results in the relation between   old and new geodesics   $\dot{\tilde{\gamma}}^\alpha=e^{-2\psi} \dot{\gamma}^\alpha$. Then, the equations of motion in the new frame read
\begin{align}
\dot{\tilde{\gamma}}^\alpha\tilde{\LCM}_\alpha \dot{\tilde{\gamma}}^\beta=0, \quad \tilde{g}_{\alpha\beta} \dot{\tilde{\gamma}}^\alpha\dot{\tilde{\gamma}}^\beta=-\tilde{m}^2,
\end{align}
where $\tilde{\LCM}_\alpha$ is a Levi-Civita connection for the metric tensor $\tilde{g}_{\alpha\beta}$.

Thus, we have reduced the motion of particles with the coordinate-dependent mass to the motion of constant mass particles in the Weyl transformed metric. However, the new metric $\tilde{g}_{\alpha\beta}$ may not have the same set of symmetries as the original one. On one hand, we can require that the new metric $\tilde{g}_{\alpha\beta}$ possesses the required symmetries independently, without any direct reference to the original metric ${g}_{\alpha\beta}$. On the other hand, if the original metric ${g}_{\alpha\beta}$ already exhibits certain symmetries, the mass function $m(x)$ must somehow share these same symmetries. Therefore, for correct application of the previously obtained results, it is necessary to present a number of requirements on the distribution of mass.

First, an arbitrary Weyl transformation $\psi$, preserves the exact Killing vectors $\kappa_a{}^\alpha$ if and only if $\kappa_a{}^\alpha\LCM_\alpha \psi=0$ which implies $\kappa_a{}^\alpha\LCM_\alpha m=0$. If the original metric is axi-stationary, then the mass function $m(x)$ must be axi-stationary too. In this case, integrals of motion coincide 
\begin{align}
\tilde{q}_a=\kappa_a{}^\alpha \tilde{g}_{\alpha\beta} \dot{\tilde{\gamma}}^\alpha=\kappa_a{}^\alpha g_{\alpha\beta} \dot{\gamma}^\alpha=q_a.
\end{align}
Second, conformal Killing tensors are preserved by Weyl transformations, but exact Killing tensors become conformal Ref. \cite{Kobialko:2022ozq}. If we want the transformed metric $\tilde{g}_{\alpha\beta}$ to posses an exact Killing tensor of rank two, then the original one must posses at least a conformal Killing tensor of rank two. Since we have used an assumption of slice-reducibility of the Killing tensor for the shadow description, we have to analyze the case of slice-reducible both conformal and exact Killing tensors. To find general condition on Weyl transformation that allows for a slice-reducible exact Killing tensor, we write down the condition from Ref. \cite{Kobialko:2022ozq}
\begin{align}
\LCS_\gamma (\tilde{\chi}_\tau\tilde{\varphi}^3) = 0.
\end{align}
Substituting the quantities associated with the original metric tensor and the Weyl transformation $\psi=\ln m/\tilde{m}$, the condition will read:
\begin{align}
    \LCS_\gamma (\tilde{\chi}_\tau\tilde{\varphi}^3)
    &=
    \LCS_\gamma \left(\varphi^3e^{2\psi}(\chi_\tau+n^\alpha\LCM_\alpha\psi)\right)
    \nonumber\\&=
    \frac{1}{2}\LCS_\gamma \left(\varphi^3e^{-\Psi}n^\alpha\LCM_\alpha (e^\Psi (m/\tilde{m})^2)\right)
    \nonumber\\&=
    \frac{1}{2}\LCS_\gamma(\varphi^2e^{-\Psi})\left(\varphi n^\alpha\LCM_\alpha (e^\Psi (m/\tilde{m})^2)\right)
    +\frac{1}{2}\varphi^2e^{-\Psi}\LCS_\gamma\left(\varphi n^\alpha\LCM_\alpha (e^\Psi (m/\tilde{m})^2)\right)\nonumber,
\end{align}
In the first transition we used the following relations \cite{Kobialko:2022ozq}
\begin{equation}
    \tilde{\varphi} = e^\psi \varphi,\qquad
    \tilde{\chi}_\tau = e^\psi\left(
        \tilde{\chi}_\tau + n^\alpha \nabla_\alpha \psi
    \right),
\end{equation}
while in the second transition we used the expression for $\psi$ and Eq. (\ref{eq:psi}) which is fair for a conformal tensor as well \cite{Kobialko:2022ozq}.
The first term is zero $\LCS_\gamma(\varphi^2e^{-\Psi})=\varphi^2e^{-\Psi}\{\LCS_\gamma\ln\varphi^2-\LCS_\gamma \Psi\}=0$ due to Eq. (\ref{eq:psi}) again. The remaining term gives us a condition of the existence of an exact Killing tensor of rank two:
\begin{align}\label{eq:m_integrability}
    \LCS_\gamma\left[\left( e^{\Psi}m^2\right)'\right]=0 \qquad \text{or}\qquad
    \LCS_\gamma\left[m^2\varphi^3 \chi_\tau + \frac{1}{2}\varphi^2\,\left(m^2\right)'\right]=0.
\end{align}
Since we assumed the existence of a slice-reducible conformal Killing tensor in the original metric, the integrability conditions (\ref{eq:integrability_1}), (\ref{eq:integrability_2}) hold automatically. Since we did not use the condition for the existence of an exact tensor in the original metric, but only a conformal one, we can consider the motion of photons in plasma and particles of variable mass even for systems without an exact tensor, if the distribution $m^2$ has a suitable form. However, if the original metric  already has an exact Killing tensor, the term $\LCS_\gamma\left(\varphi^3 \chi_\tau\right)$ is zero, and the mass function must obey the condition $\LCS_\gamma (m^2)\varphi^3 \chi_\tau+\LCS_\gamma\left(\varphi^2\,  \left(m^2\right)' \right)=0$.

Generalization of Eq. (\ref{eq:result_1}) can be performed by replacing all the quantities associated with the original spacetime with the metric $g_{\alpha\beta}$ by the new quantities associated with the transformed spacetime $\tilde{g}_{\alpha\beta}$ and expressing them back in terms of the quantities with no tilde. Recall that according to Ref. \cite{Kobialko:2022ozq} the quantity $e^{\Psi}\G^{ab}$ is invariant and  
\begin{align}
\tilde{\G}^{\alpha\beta}=e^{-2\psi}\G^{\alpha\beta},\quad
\tilde{\bar{v}}^a&=\bar{v}^a, \quad \tilde{\bar{v}}=e^{\bar{\psi}}\bar{v}, \quad \tilde{\bar{\tau}}^a= \bar{\tau}^a, \quad 
\tilde{\bar{E}} = \bar{E}, \quad \tilde{m}_E=\bar{m}_E, \quad \tilde{\Psi}=2\psi +\Psi, 
\end{align}
and
\begin{align}
\tilde{\mathcal{S}}^{ab}= e^\Psi\{\G^{ab} + e^\Psi e^{2(\psi-\bar{\psi})} \bar{m}^2_E\cdot\bar{v}^a\bar{v}^b/\bar{v}^2\}=e^\Psi\{\G^{ab} + e^\Psi  m^2_E\cdot\bar{v}^a\bar{v}^b/\bar{v}^2\}, \quad \tilde{\bar{v}} e^{-\tilde{\bar{\Psi}}/2}=\bar{v} e^{-\bar{\Psi}/2},
\end{align}  
where $m^2_E=\bar{v}^2 m^2/ \bar{E}^2$, $\bar{m}^2_E=\bar{v}^2 \bar{m}^2/ \bar{E}^2$ are effective masses at the observer's and massive particle surface's point (lying on the same integral curve), respectively. As a result of these transformations, the only modification we have to do when we move from the system with constant mass to the system with variable mass is just considering $m^2_E$ as variable in shadow matrix $\mathcal{S}^{ab}$ and considering constant $\bar{m}^2_E$ in $\overline{\mathcal{S}}{}^{ab}$ and $\sqrt{1-\bar{m}^2_E}$. Moreover, Eq. (\ref{eq:xi_sol}) also remains unchanged, with the caveat that the mass is variable now and it must be differentiated. Since we do not change the foliation, the derivative is also invariant, from Eq. (\ref{eq:m_integrability}) follows the unchanged form of the condition in Eq. (\ref{eq:m_integrability_1})
\begin{align}\label{eq:m_integrability_2}
    \LCS_\gamma\left[\left(\mathcal{S}^{ab}\right)'\right]=0.
\end{align}

\subsection{General result and discussion}

Summarizing the entire procedure for construction of a gravitational shadow for massive particles (probably, with variable mass) in spacetimes with slice-reducible Killing tensor of rank two, one can highlight three steps:
\begin{enumerate}
    \item Find compact slices $S$ generating the slice-reducible exact Killing tensor (Ref. \cite{Kobialko:2022ozq}). Calculate quantities $n^\alpha$, $\Psi$, $\G_{ab}$ for the slices. If the particles under consideration have variable mass, the Killing tensor can be conformal and not exact, with two additional condition imposed on the mass function: $\LCS_\alpha\left(\left( e^{\Psi}m^2\right)'\right)=0$ and $\kappa_a{}^\alpha\LCM_\alpha m=0$.
    \item Define an observer at some point $O$ with the four-velocity vector $\bar{v}^a$ and fix the particle energy $\bar{E}$ detected by the observer. Construct an integral curve of the normal field $n^\alpha$ passing through the observation point.
    \item Calculate $X$ and $Y$ for each slice at the  point related to the observer through the integral curve using formulas (or Eq. (\ref{eq:Theta_Phi})):
    \begin{subequations}\label{eq:XY_general_all}
    \begin{align} \label{eq:XY_general}
    X &=
     \frac{
        2\bar{\tau}^a\rho_a
    }{
       \sqrt{1-\bar{m}^2_E}\cdot\bar{v}^a\rho_a
        \mp \bar{v} e^{-\bar{\Psi}/2} \sqrt{\left(\mathcal{S}^{ab}-\overline{\mathcal{S}}{}^{ab}\right)\rho_a\rho_b
        }
    }, \\
    Y &=
      \pm \frac{
        2 \bar{v} e^{-\bar{\Psi}/2}\sqrt{ -\mathcal{S}^{ab}\rho_a\rho_b }
    }{
      \sqrt{1-\bar{m}^2_E}\cdot\bar{v}^a\rho_a
        \mp \bar{v} e^{-\bar{\Psi}/2} \sqrt{\left(\mathcal{S}^{ab}-\overline{\mathcal{S}}{}^{ab}\right)\rho_a\rho_b}
    },
   \end{align}
   where $\rho_a$ is an arbitrary non-trivial solution to the equation
   \begin{align} \label{eq:PbarP}
   \left(\mathcal{S}^{ab}\right)'\rho_a\rho_b=0, \quad \mathcal{S}^{ab}= e^\Psi \left(\G^{ab} + m^2_E\cdot\bar{v}^a\bar{v}^b/\bar{v}^2 \right), \quad
   m^2_E=\bar{v}^2 m^2/ \bar{E}^2,
   \end{align}
   \end{subequations}
   and the prime $'$ means the derivative with respect to the foliation parameter $\varphi n^\alpha \LCM_\alpha$ (keep in mind that the observer's coordinate are not differentiated). The solution $\rho_a$ must satisfy Eqs. (\ref{eq:rho_c}) and (\ref{eq:MPR}), otherwise $X$ and $Y$ become imaginary. Coordinates $X$ and $Y$ of the boundary are parameterized as a function of the foliation parameter $\Omega$. The resulting curve should be analyzed to determine whether it corresponds to the shadow boundary or the boundary of relativistic images (further details will be provided in examples of  Section \ref{sec:examples}).
   \end{enumerate}

There are several distinctive features related to the choice of signs in the expressions (\ref{eq:XY_general_all}):
\begin{itemize}
    \item The sign $\mp$ in the denominator refers to different stereographic projections of the shadow image related by inversion  $\rho_a\rightarrow-\rho_a$. For the future-directed case $\rho_a\bar{v}^a<0$ in most cases we can choose the sign $-$. Moreover, Eq. (\ref{eq:XY_general_all}) is invariant under transformations $\rho_a\rightarrow s\rho_a$ for some arbitrary $s>0$.
    \item The general sign $\pm$ in front of the fraction $Y$ is independent and provides a mirror symmetry of the shadow with respect to the line $Y=0$ \cite{Perlick:2021aok}. While this symmetry is expected when the observer is located at the equatorial plane, it seems counter-intuitive when the observer is off-plane. However, the mirror symmetry of the shadow for any observers arises due to the independence of the Killing tensor from the choice of the direction tangent vector $\bar{e}_2{}^\alpha$ (e.g., $\bar{e}_2{}^\alpha \sim \partial_\theta$ in Kerr spacetime). So, considering the geodesic curve with $\dot{\gamma}_\alpha \bar{e}_2{}^\alpha \to -\dot{\gamma}_\alpha \bar{e}_2{}^\alpha$ leads to the same integrals of motion and the same massive particle surface, where these geodesics wind up.
    \item The expression for shadow also has the symmetry $e^\Psi\rightarrow \text{const} \cdot e^\Psi$. This symmetry is associated with the freedom to multiply the Killing tensor by an arbitrary positive constant. 
    \item Another feature is that the Killing tensor (\ref{eq:KT}) does not contain terms linear in the normal vector $n^\alpha$. This leads to appearance of the symmetry $n^\alpha\rightarrow - n^\alpha$ in all obtained expressions. As a result, we can unambiguously determine the image of a shadow only on the projective celestial sphere, in which the opposite points are identified. In order to obtain a real image of the shadow, it is necessary to discard an extra piece of the image using additional physical considerations. For example, for a far away observer, the shadow will be located in the vicinity of the zenith (if the normal and the tetrad are chosen as described above, i.e., the observer's camera is pointing towards the gravitating object). So, we must choose the $+$ sign of $\cos \Theta$, resolving the ambiguity.
\end{itemize}

As a result of applying Eq. (\ref{eq:XY_general}), we get the curve describing the boundary of the relativistic images \cite{Virbhadra:1999nm,Tsupko:2013cqa} parameterized by the foliation parameter $\Omega$. For black hole objects, this corresponds to the boundary of the shadow. The described procedure is completely independent on the coordinate choice, though a special choice may be more convenient. Moreover, the task of solving geodesic equations is no longer necessary and the construction of the final formula comes down to simple calculations. Of course, the main problem is now concentrated in the construction of the slice-reducible Killing tensor and the corresponding slices. However, this problem was formulated in a coordinate-independent way in Refs. \cite{Kobialko:2022ozq,Kobialko:2021aqg}. The expression for the boundary itself is simpler and explicitly coordinate independent than, for example, in Refs. \cite{Perlick:2017fio,Briozzo:2022mgg,Perlick:2023znh,Bezdekova:2022gib,Atamurotov:2015nra,Bisnovatyi-Kogan:2017kii,Badia:2021kpk} and can be studied analytically from various points of view. In particular, the properties of the shadow matrix are closely related to the geometric properties of massive particle surfaces. Similarly to the photon surfaces \cite{Shiromizu:2017ego,Feng:2019zzn,Yang:2019zcn}, massive particle surfaces are subject to various geometric restrictions. These restrictions can ultimately lead to rather universal restrictions on the shadows parameters in  presence of plasma \cite{Feng:2019zzn,Crisnejo:2018uyn,Yang:2019zcn}.  

There is one more useful property of expressions in Eq. (\ref{eq:XY_general_all}). Namely, despite the fact that the expressions are not tensors indexed by $a$, $b$, they are invariant under transformation of the basis in the space of Killing vectors with constant coefficients ${\kappa_{a'}{}^\alpha=\Lambda^a_{a'}\kappa_a{}^\alpha}$. Indeed, if the transformation matrix $\Lambda^a_{a'}$ is constant, all contractions are invariant, including those with derivatives (e.g., $\left(\mathcal{S}^{ab}\right)'\rho_{a}\rho_{b}$), or involving vectors from different points (e.g., $\overline{\mathcal{S}}{}^{ab} \rho_{a}\rho_{b}$). Expressions (\ref{eq:XY_general_all}) are not invariant under a more general transformation, since we calculate contractions of tensors and vectors (of the Killing vector space) at different points of spacetime. In general, such contractions are not valid. In our case it is only possible because we can transfer vectors using combinations of Killing vectors with constant coefficients along the integral curves of the foliation normals.  
As an exception, we can use a transformation that does not depend on the foliation parameter, $\left(\Lambda^a_{a'}\right)'=0$, but in this case we lose the property from Eq. (\ref{eq:m_integrability_2}).

\subsection{Shadows for distant observers in asymptotically flat spacetimes}

As a rule, we are interested in shadows that are seen by an observer far away from the gravitating object, i.e., the distance from the observer to the gravitating object is much larger than the gravitational radius of the object. 
In this case, it is reasonable to consider the observer asymptotically distant. At large distances, the gravitating object is seen as a point source of mass with all higher multipoles suppressed. Asymptotically, slices tend to be convex spheres with $\chi_\tau > 0$. The parameter measuring the distance from the object is chosen in the form of a foliation parameter $\Omega$, tending to infinity for the slice where the asymptotically distant observer $\bar{\Omega}\rightarrow \infty$ lives (here $\bar {\Omega}$ — $\Omega$ value for a slice with an observer).
In many particular examples, the parameter $\Omega$ coincide with the Boyer-Lindquist radius $r$. Though, the situation can be less obvious for solutions with the gravimagnetic mass (NUT parameter) and causality violation, the following calculations are fair for them as well. In what follows, we assume for simplicity that the coordinate systems tends asymptotically to spherical coordinate system with two Killing vectors $\partial_t$ and $\partial_\phi$, though similar steps can be applyed to more general spacetimes and coordinates.


Let us pick out bounded and unbounded parts of this limit. The scalar $\Psi$ grows unbounded due to the equation: $n^\alpha \LCM_\alpha \Psi=2\SFS_\tau$. Indeed, for asymptotically spherical surfaces, the quantity associated with the curvature $\chi_\tau$ is positive and decreases as $\sim 1/r$, where $r$ is a radius in a coordinate system, which tends to the spherical one. Thus, we expect logarithmic growth: $\Psi \sim \ln r$.

We expect that the observer has a finite speed $\bar{v}$, assuming that vector components $\bar{v}^a$ are bounded either. This allows us to define the following limits
\begin{align}
\bar{v}^a_\infty\equiv \lim_{\bar{\Omega}\rightarrow\infty}(\bar{v}^a), \quad \bar{v}_\infty\equiv \lim_{\bar{\Omega}\rightarrow\infty}(\bar{v}).
\end{align}
The effective particle mass   $\bar{m}$ must be also bounded 
\begin{align}
\bar{m}_\infty\equiv \lim_{\bar{\Omega}\rightarrow\infty}(\bar{m}), \quad m^2_{E\infty}=\bar{v}^2_\infty m^2/ \bar{E}^2, \quad \bar{m}^2_{E\infty}=\bar{v}^2_\infty \bar{m}^2_\infty / \bar{E}^2,
\end{align}
or even zero $\bar{m}_\infty=0$ if we consider photons in plasma. 
Then, the shadow matrix occurs to be finite:
\begin{align}
 \mathcal{S}^{ab}_\infty \equiv \lim_{\bar{\Omega}\rightarrow\infty}( \mathcal{S}^{ab})&=e^\Psi \left(\G^{ab} + m^2_{E\infty} \bar{v}^a_\infty \bar{v}^b_\infty/\bar{v}^{2}_\infty \right). 
\end{align}
The solution $\rho^\infty_a$ defined by equation 
\begin{align}   \label{eq:rho_infty}
 \left(\mathcal{S}^{ab}_\infty\right)'\rho^\infty_a\rho^\infty_b=0, 
\end{align}
can also be considered finite. Though, the observer with non-zero finite linear velocity can have infinite angular momentum, we will not consider this case since it introduces well-known aberrations that can be obtained in Special Relativity framework (as a rule, the effect is called light aberration, but in our case this effect should be considered for massive particles \cite{Briozzo:2022mgg}):
\begin{align}
\bar{v}_a^\infty\equiv\lim_{\bar{\Omega}\rightarrow\infty}(\bar{\G}_{ab}\bar{v}^b).
\end{align}
Static and ZAMO observers have finite $\bar{v}_a^\infty$. Particularly, one can consider a limiting procedure such that the angular momentum can approach any finite value, though the linear velocity in the azimuthal direction is zero.

However, vector components $\bar{\tau}^a$ have $\det\bar{\G}_{ab}$ in their definition, which can be not finite, e.g., for asymptotically spherical coordinate system. Thus, it is reasonable to pick out an explicitly finite part of the vector components as follows:
\begin{align}
\bar{\tau}^{a }_\infty\equiv  \lim_{\bar{\Omega}\rightarrow\infty}(\bar{\tau}^a\sqrt{-\det\bar{\G}_{ab}})=\epsilon^{ab}\bar{v}_a^\infty.
\end{align} 
In particular, from Eq. (\ref{eq:fgr}) follows a non-trivial asymptotic behavior
\begin{align} 
\overline{\mathcal{S}}{}^{ab}\rho^\infty_a\rho^\infty_b\approx -e^{\bar{\Psi}}(1-\bar{m}^2_E)\cdot(\bar{v}^c_\infty\rho^\infty_c)^2/\bar{v}^{2}_\infty+e^{\bar{\Psi}}/\sqrt{-\det\bar{\G}_{ab}}\cdot(\bar{\tau}^a_\infty\rho^\infty_a)^2/\bar{v}^{2}_\infty, 
\end{align}
Substituting all expressions back into Eq. (\ref{eq:XY_general_all}), and keeping only the first non-zero term of the expansion with respect to $\bar{\Omega} \to \infty$, we find (we retain only the one sign $-$ in the denominator for future directed directions $\rho_a\bar{v}^a<0$, discarding the mirror image on the projective sphere)
\begin{align} \label{eq:xy_2}
X\approx  \frac{A(\Omega)}{\sqrt{-\det\bar{\G}_{ab}}}, \quad Y\approx\pm B(\Omega) e^{-\bar{\Psi}/2},
\end{align}
where $\Omega$ is a finite foliation parameter of the massive particle surface. The observer is placed infinitely far away from the gravitating object, so the shadow image must be infinitely small. Since any aberrations are absent due to our choice of the observer, the image must be placed right in the zenith without any shifts. This conclusion is confirmed by Eq. (\ref{eq:xy_2}) if we take into account that $\sqrt{-\det\bar{\G}_{ab}}$ and $e^{\bar{\Psi}/2}$ tend to infinity. In particular, condition (\ref{eq:rho_c}) reduces to $\rho^\infty_a \bar{v}^a_\infty\neq0$. Multiplying $X$ and $Y$ by $e^{\bar{\Psi}/2}$ results in
\begin{subequations}\label{eq:XY_infty}
\begin{align} 
    X_\infty &\equiv \lim_{\bar{\Omega}\rightarrow\infty}(Xe^{\bar{\Psi}/2})=
    \frac{\alpha_\infty \cdot(\bar{\tau}^a_\infty\rho^\infty_a)
    }{
       \sqrt{1-\bar{m}^2_{E\infty }}\cdot (\bar{v}^a_\infty\rho^\infty_a)
    },\\
    Y_\infty &\equiv \lim_{\bar{\Omega}\rightarrow\infty}(Ye^{\bar{\Psi}/2})=
    \pm \frac{ \bar{v}_\infty\cdot\sqrt{-\mathcal{S}^{ab}_\infty\rho^\infty_a\rho^\infty_b} }{
        \sqrt{1-\bar{m}^2_{E\infty }}\cdot(\bar{v}^a_\infty\rho^\infty_a)
    },
\end{align}
where
\begin{align}
\alpha_\infty\equiv\lim_{\bar{\Omega}\rightarrow\infty}(e^{\bar{\Psi}/2}/\sqrt{-\det\bar{\G}_{ab}}), \quad \left(\mathcal{S}^{ab}_\infty\right)'\rho^\infty_a\rho^\infty_b=0.
\end{align}
\end{subequations}
Note that the function $\Psi$ is defined up to some additive constant $\Psi \to \Psi + 2C$, which can be used to control a scale of the image: $(X_\infty,Y_\infty)\rightarrow e^C (X_\infty,Y_\infty)$. If $e^\Psi m^2_{E\infty}$ can be neglected in $\mathcal{S}^{ab}_\infty$, coordinates $X_\infty$, $Y_\infty$ scales as
\begin{equation}\label{eq:scale_approximation}
    (X_\infty, Y_\infty) \Big|_{\bar{m}^2_{E\infty }} \approx \frac{(X_\infty, Y_\infty) \Big|_0}{\sqrt{1-\bar{m}^2_{E\infty }}}.
\end{equation}

\section{Explicit form and Examples}
\label{sec:examples}

\subsection{Benenti-Francaviglia form}

All results obtained above were obtained for an arbitrary coordinate system, and instead of an explicit search for a separable coordinate system, it is required to know the corresponding foliation. However, an explicit relation between the separable coordinate system and the foliation is presented in Ref. \cite{Kobialko:2022ozq}. Adopting the result for our case, it was shown that if a slice-reducible conformal Killing tensor of rank two and two commuting Killing vectors exist in a four-dimensional spacetime, then there is a coordinate system with metric tensor of the form (see discussion of the Benenti-Francaviglia ansatz in Refs. \cite{Benenti:1979erw,Papadopoulos:2018nvd,Carson:2020dez,Papadopoulos:2020kxu})
\begin{equation} \label{eq:BFF}
    ds^2 = 
    \lambda(r,\theta)
    \left[
        (\mathcal{F}^{-1})_{ab}dy^a dy^b + f_r(r) dr^2 + f_\theta(\theta) d\theta^2
    \right],
\end{equation}
where matrix $\mathcal{F}^{ab}(r, \theta) = \mathcal{X}_r^{ab}(r) + \mathcal{X}_\theta^{ab}(\theta)$ is separable, functions $f_r(r)$, $f_\theta(\theta)$, $\mathcal{X}^{ab}_r(r)$, $\mathcal{X}^{ab}_\theta(\theta)$, $\lambda(r,\theta)$ are arbitrary. We denoted coordinates with letters $r$ and $\theta$ to stay connected with common metrics, but they can be arbitrary in general. Vectors along coordinates $y^a$ represent Killing vectors, usually, representing a timelike vector $\partial_t$ and azimuthal spacelike Killing vector $\partial_\phi$. The foliation slices are determined by $r=\text{const}$ with foliation parameter $\Omega = r$. The corresponding conformal Killing tensor is
\begin{align} \label{SMK}
   & K^{\alpha\beta} =
    \alpha(r,\theta) g^{\alpha\beta}
    + \mathcal{X}_r^{ab}(r) \delta_{a}{}^\alpha \delta_{b}{}^\beta
    +f_r(r)^{-1}\delta^{\alpha}_{r} \delta^{\beta}_{r}.
\end{align} 
The geometric quantities of this foliation are
\begin{subequations}
\begin{equation}
   \varphi n^\alpha \nabla_\alpha = \partial_r,\quad
    \chi_\tau = \frac{1}{2} (\lambda f_r)^{-1/2} \partial_r \ln (\lambda f_\theta),\quad
    \varphi = (\lambda f_r)^{1/2},
\end{equation}
\begin{equation} 
\Psi=\ln \lambda(r,\theta), \quad 
\G_{ab} = \lambda \left(\mathcal{F}^{-1}\right)_{ab}.
\end{equation} 
\end{subequations}
Additionally, if we are interested in exact slice-reducible Killing tensors, the integrability condition $\LCS_\gamma (\chi_\tau\varphi^3) = 0$ gives $\partial_\theta \partial_r \lambda = 0$, thus the conformal factor $\lambda$ must be a function of the form $\lambda = \mathcal{X}_r^{\lambda}(r) + \mathcal{X}_\theta^{\lambda}(\theta)$ resulting in $\alpha = - \mathcal{X}_r^\lambda(r)$ (for comparison see Refs. \cite{Papadopoulos:2018nvd,Carson:2020dez,Papadopoulos:2020kxu}, where a similar form of the metric was proposed from other considerations). In the case of massive particles, we have to consider only exact type of Killing tensors. However, the presence of plasma of a suitable type allows one to consider conformal tensors. In this case, following Eq. (\ref{eq:m_integrability}), we must have (see Refs. \cite{Bezdekova:2022gib,Perlick:2017fio,Perlick:2023znh})
\begin{equation} 
\lambda m^2_E=\M^E(r,\theta)= \M^E_r(r)+\M^E_\theta(\theta).
\end{equation} 
The shadow matrix reads
\begin{equation} 
\mathcal{S}^{ab} = \F^{ab} +\M^E \cdot\bar{v}^a\bar{v}^b/\bar{v}^2,
\end{equation} 
where observer's velocity $\bar{v}^a$ and its dual vector $\bar{\tau}^a$ read
\begin{equation} 
\bar{v}^a = \begin{pmatrix}
   \bar{v}^t\\
    \bar{v}^\phi
    \end{pmatrix}, \quad \bar{\tau}^a =\frac{\bar{\F}^{ab}\epsilon_{bc}\bar{v}^c}{\sqrt{-\det \bar{\F}^{ab}}}, \quad  \bar{v}=\sqrt{-\bar{\lambda} (\bar{\mathcal{F}}^{-1})_{ab}\bar{v}^a\bar{v}^b}.
\end{equation} 
To find $\rho_a$ explicitly, the shadow matrix $\mathcal{S}^{ab}$ can be expanded into a Killing basis with the following norm fixation
\begin{equation} \label{eq:rho_norm}
\rho_a{} = \begin{pmatrix}
    -1\\
    \rho_\phi
    \end{pmatrix}.
\end{equation} 
As a consequence of Eq. (\ref{eq:q_sol}), in this norm we have $\rho_\phi=- q_\phi / q_t$, i.e. $\rho_\phi$ is an ordinary impact parameter of the geodesic. From Eq. (\ref{eq:PbarP}), we obtain a quadratic equation, the solution to which is:
\begin{equation} 
\rho^\pm_\phi =\frac{\left(\mathcal{S}^{t\phi}\right)'\mp\sqrt{ - \det \{ \left(\mathcal{S}^{ab}\right)'\}}}{\left(\mathcal{S}^{\phi\phi}\right)'}, \quad \det \{ \left(\mathcal{S}^{ab}\right)'\}=\left(\mathcal{S}^{tt}\right)'\left(\mathcal{S}^{\phi\phi}\right)'-\left(\mathcal{S}^{t\phi}\right)'\left(\mathcal{S}^{t\phi}\right)'.
\end{equation} 
We get two solutions $\rho^+_\phi$, $\rho^-_\phi$ and in general we should hold both of them. However, for numerous examples, one of the solutions may not correspond to any massive particle region, yielding non-real values for $X$ and $Y$, and should therefore be discarded (see Ref. \cite{Perlick:2023znh} for details). The formula for the boundary is obtained by directly substituting these expressions into (\ref{eq:XY_general_all}), and using Eq. (\ref{eq:fgr}). After some simplifications we get
\begin{subequations} \label{eq:XY_BF}
    \begin{align}
    X &=
     \frac{
        2\bar{\tau}^a\rho_a
    }{
       \sqrt{1-\bar{m}^2_E}\bar{v}^a\rho_a
        \mp \sqrt{(1-\bar{m}^2_E)(\bar{v}^c\rho_c)^2-(\bar{\tau}^a\rho_a)^2+\frac{\bar{v}^{2}}{\bar{\lambda}}\cdot\mathcal{S}^{ab} \rho_a \rho_b
    }
    }, \\
    Y &=
      \pm\frac{
         2 \sqrt{-\frac{\bar{v}^{2}}{\bar{\lambda}}\cdot\mathcal{S}^{ab} \rho_a \rho_b}
    }{
      \sqrt{1-\bar{m}^2_E}\bar{v}^a\rho_a
        \mp \sqrt{(1-\bar{m}^2_E)(\bar{v}^c\rho_c)^2-(\bar{\tau}^a\rho_a)^2+\frac{\bar{v}^{2}}{\bar{\lambda}}\cdot\mathcal{S}^{ab} \rho_a \rho_b}
    },
   \end{align}
where
\begin{align}
\mathcal{S}^{ab} \rho_a \rho_b=- \frac{\left(\mathcal{S}^{\phi\phi}\right)^2}{ \left(\mathcal{S}^{\phi\phi}\right)'}
    \left[
     \left(\frac{\mathcal{S}^{tt}}{\mathcal{S}^{\phi\phi}}\right)'
     -2
     \left(\frac{\mathcal{S}^{t\phi}}{\mathcal{S}^{\phi\phi}}\right)'
     \rho^{\pm}_{\phi}
     \right].
\end{align}
\end{subequations} 
Eqs. (\ref{eq:XY_BF}) represent a general expression for the shadow boundary of an arbitrary metric that can be written in Benenti-Francaviglia form or its conformal generalization. This family includes the Plebanski-Demianski solution \cite{Stephani:2003tm,Demianski:1980mgt}, EMD \cite{Rasheed:1995zv}, EMDA \cite{Galtsov:1994pd}, STU rotating black holes \cite{Chow:2014cca}. In asymptotically spherical coordinates, the asymptotics of the functions are as follows:
\begin{align} 
\G_{ab}&
\to
\begin{pmatrix}
- 1 & -N_\infty(\theta)\\
-N_\infty(\theta) &  r^2 \sin^2\theta
\end{pmatrix},
\quad
\lambda(r,\theta)f_r(r)\to1, \quad
\lambda(r,\theta)f_\theta(\theta)  \to r^2,
\end{align}
where $N_\infty(\theta) = 2N(\cos\theta + C_{N})$, $N$ is the NUT parameter, and $C_N$ is a constant for the NUT gauge. This asymptotic implies that $f_\theta\to C$, $f_r \to C / r^2$, $\lambda \to r^2/C$ for some unimportant constant $C$, which is natural to choose equal to $1$.

Let us asymptotically expand quantities related to the stationary observer at the point $(\bar{r}, \bar{\theta})$. The requirement that the speed $\bar{v}$ and the observer’s conserved quantities $\bar{v}_a$ are bounded leads to the following general expression (static, ZAMO) 
\begin{align}
    \bar{v}^a&=
    \begin{pmatrix}
    1\\
    w_\infty/\bar{r}^2
    \end{pmatrix},
\end{align}
where $w_\infty$ is a constant and the length of the vector is chosen to be equal to unity $\bar{v} = 1$ for simplicity, since it does not influence the final result. This gives us the following limits 
\begin{align}
\bar{v}^a_\infty=\begin{pmatrix} 1 \\
0
\end{pmatrix}, \quad \bar{v}_a^\infty=\begin{pmatrix} -1 \\
W
\end{pmatrix},\quad
\bar{\tau}^a_\infty=\begin{pmatrix}  - W \\
-1
\end{pmatrix},\quad
    \alpha_\infty=\sin^{-1}\bar{\theta},
\end{align}
where $W = w_\infty\sin^2\bar{\theta}-N_\infty$. Then from (\ref{eq:rho_infty}) for $\rho^{\infty}_{a}$ in the same norm fixation (\ref{eq:rho_norm}) we find
\begin{align} \label{eq:rho_EF}
 \rho^{\infty\pm}_{\phi}=
\frac{ \left(\F^{t\phi}\right)'\mp\sqrt{\left(\F^{t\phi}\right)'^2-\left(\F^{tt}+\M^E\right)' \left(\F^{\phi\phi}\right)'}}{ \left( \F^{\phi\phi}\right)'}.
\end{align}
By collecting everything together and simplifying the expressions, we find
\begin{subequations} \label{eq:XY_asymptotic}
\begin{align} 
    X_\infty &=\lim_{r\rightarrow\infty}(Xr)=
    \frac{
       \rho^{\infty\pm}_{\phi}-W
    }{
        \sin\bar{\theta}\sqrt{1-\bar{m}^2_{E\infty }}
    },
    \quad
    Y_\infty = \lim_{r\rightarrow\infty}(Yr)=
    \mp \frac{ \sqrt{ -\mathcal{S}^{ab}_\infty \rho^\infty_a \rho^\infty_b}}{
        \sqrt{1-\bar{m}^2_{E\infty }}
    },
\end{align}
where
\begin{align}
\mathcal{S}^{ab}_\infty \rho^\infty_a \rho^\infty_b=-\frac{\left(\F^{\phi\phi}\right)^2}{ \left(\F^{\phi\phi}\right)'}
    \left[
     \left(\frac{\F^{tt}+\M^E}{\F^{\phi\phi}}\right)'
     -2
     \left(\frac{\F^{t\phi}}{\F^{\phi\phi}}\right)'
     \rho^{\infty\pm}_{\phi}
     \right], \quad \bar{m}^2_{E\infty }=\lim_{r\rightarrow\infty}(\lambda^{-1}\M^E).
\end{align}
\end{subequations}
Note that changing $w_\infty$ leads to the usual shift of the shadow along axis $X$ since the shadow matrix is independent of $w_\infty$ in asymptotic limit.

\subsection{Boyer-Lindquist coordinates}

Let us commence with the simplest example of the Kerr metric and similar solutions, including EMD, EMDA, and STU black holes \cite{Stephani:2003tm,Demianski:1980mgt,Rasheed:1995zv,Galtsov:1994pd,Chow:2014cca}. In Boyer-Lindquist coordinates \cite{Stephani:2003tm}, the metric assumes the following straightforward form
\begin{equation} \label{SolK}
    ds^2 =
    - \frac{\Delta - a^2 \sin^2\theta}{\Sigma} \left(dt - \omega d\phi\right)^2
    + \Sigma \left(
        \frac{dr^2}{\Delta} + d\theta^2 + \frac{\Delta \sin^2\theta}{\Delta - a^2 \sin^2\theta} d\phi^2
    \right).
\end{equation}
It is well-known that these solutions possess at least conformal Killing tensors reducible on slices $r=\text{const}$. Thus, we can represent these solutions in the form (\ref{eq:BFF}). Initially, we can readily identify the factor $\lambda=\Sigma$ and determine the matrix:
\begin{align}
\mathcal{F}^{ab}&=\frac{\Delta - a^2 \sin^2\theta}{\sin^2\theta\Delta}\begin{pmatrix}
\omega^2  & \omega \\
\omega & 1
\end{pmatrix}+\begin{pmatrix}
-\frac{\Sigma^2}{\Delta - a^2 \sin^2\theta}  & 0 \\
0 & 0
\end{pmatrix}.
\end{align}
From the existence of the Killing tensor we know that this matrix is separable, but we do not need to separate variables explicitly. Instead we need to calculate the derivative:
\begin{align}
\left(\mathcal{F}^{ab}\right)'&=\frac{a^2\Delta'}{\Delta^2}\begin{pmatrix}
\omega^2  & \omega \\
\omega & 1
\end{pmatrix}+\frac{\Delta - a^2 \sin^2\theta}{\sin^2\theta\Delta}\begin{pmatrix}
2\omega & 1 \\
1 & 0
\end{pmatrix}\omega'+\begin{pmatrix}
-\left(\frac{\Sigma^2}{\Delta-a^2 \sin ^2\theta}\right)'  & 0 \\
0 & 0
\end{pmatrix}.
\end{align}
Since the general expression (\ref{eq:XY_BF}) is rather cumbersome, we will give an explicit form for the asymptotic formula (\ref{eq:XY_asymptotic}) only. Applying Eqs. (\ref{eq:rho_EF}) and (\ref{eq:XY_asymptotic}) allows obtaining an expression of the shadow boundary for an asymptotically distant observer
\begin{subequations} \label{eq:BLXY_G}
\begin{align} \label{eq:BLXY}
X_\infty=&\frac{1
    }{
        \sin\bar{\theta}\sqrt{1-\bar{m}^2_{E\infty }}
    }\left(\omega-W +\left\{\RS\omega'\mp\sqrt{\left(\RS\omega'\right)^2+\frac{\Delta^2\mathcal{Y}'}{a^2 \Delta'  }}\right\}\right),\\
Y_\infty=&\pm\frac{ 1 }{
        \sqrt{1-\bar{m}^2_{E\infty }}
    } \left(\mathcal{Y}-\RS\mathcal{Y}'-\frac{2\Delta'a^2}{\Delta^2}\RS^2\omega'\left\{\RS\omega'\mp\sqrt{\left(\RS\omega'\right)^2+\frac{\Delta^2\mathcal{Y}'}{a^2 \Delta'  }}\right\}\right)^{1/2},
\end{align}
where 
\begin{align}
\RS\equiv\frac{\Delta - a^2 \sin^2\bar{\theta}}{a^2\sin^2\bar{\theta}}\frac{  \Delta}{\Delta' }  , \quad \mathcal{Y}\equiv\frac{\Sigma^2}{\Delta - a^2 \sin^2\bar{\theta}}-\M^E,
\end{align}
\end{subequations}
and $\bar{m}^2_{E\infty }=\bar{m}_\infty/\bar{E}$ and $\bar{m}_\infty$ is an asymptotic mass of the particle. The massive particle region  is described by the inequality
\begin{align}
\mathcal{Y}-\RS\mathcal{Y}'\geq\frac{2\Delta'a^2}{\Delta^2}\RS^2\omega'\left\{\RS\omega'\mp\sqrt{\left(\RS\omega'\right)^2+\frac{\Delta^2\mathcal{Y}'}{a^2 \Delta'  }}\right\}.
\end{align}
These formulas generalize results from Refs. \cite{Perlick:2017fio,Briozzo:2022mgg,Perlick:2023znh,Bezdekova:2022gib,Grenzebach:2014fha,Grenzebach:2015oea,grenzebach2016shadow}, since we do not specify the expression for $\omega$ explicitly, which can vary from model to model. They clearly express the contour of the shadow through the components of the metric in the usual form. 

\subsubsection{Kerr-NUT spacetime}

We begin our exploration with the vacuum Kerr-NUT spacetime, characterized by essential parameters: mass $M$, rotational Kerr parameter $a$, and the Newman-Unti-Tomburino parameter $N$ representing the gravimagnetic mass. The choice of the vacuum Kerr-NUT spacetime serves a dual purpose. Firstly, it provides a familiar foundation for evaluating the developed framework by leveraging well-known results \cite{Bisnovatyi-Kogan:2017kii,Perlick:2017fio,Badia:2021kpk}. Secondly, we illuminate the versatility and robustness of the framework introduced in this paper. This solution, denoted by the Kerr-NUT metric, assumes the form (\ref{SolK}) with the following functions:
\begin{subequations}
\begin{align}
\Delta &=r (r-2 M)+a^2-N^2,\quad \Sigma=r^2+(a \cos \theta +N)^2, \\
\omega &=-\frac{2 \left(a \sin^2\theta \left(M r+N^2\right)+\Delta N \cos \theta   \right)}{\Delta-a^2 \sin^2\theta}.
\end{align}
\end{subequations}

\begin{figure}
    \centering
    \includegraphics[width=1\linewidth]{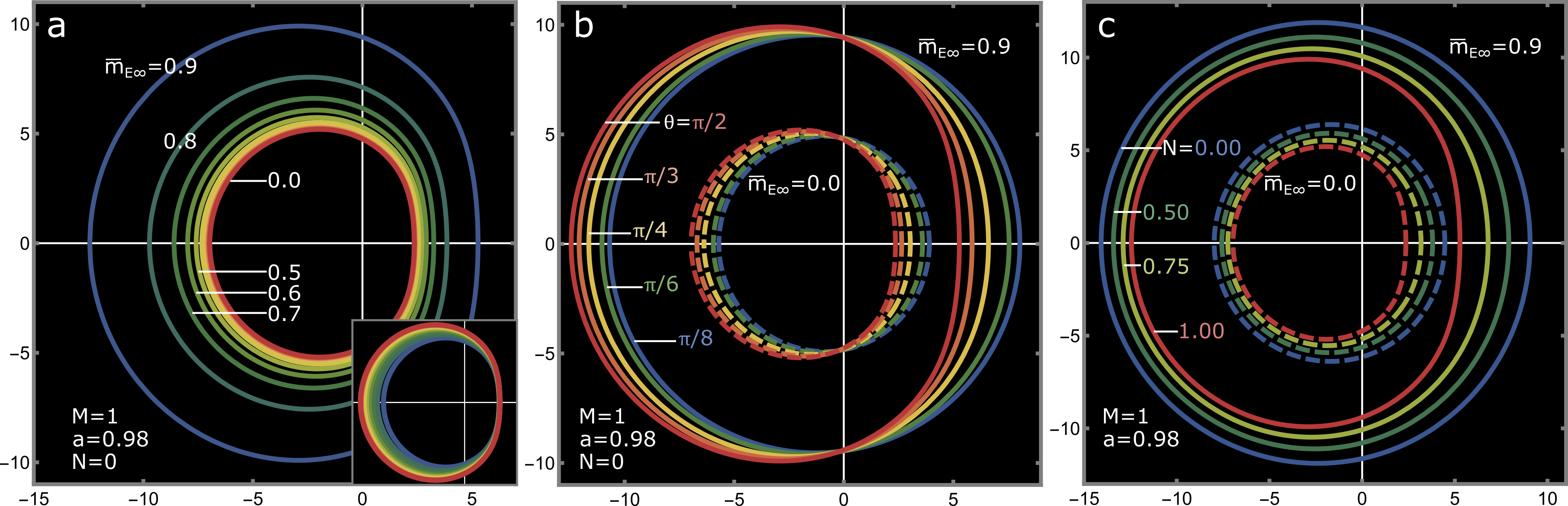}
    \caption{
    Shadows cast by particles with non-zero constant mass in the Kerr-NUT spacetime with parameters $M=1$ and $a=0.98$. (a) Illustrates shadows produced by particles with $\bar{m}_{E\infty}$ ranging from 0 to 0.9 in a spacetime with $N=0$; the inset presents the same set of shadows, rescaled by the factor $\sqrt{1-\bar{m}_{E\infty}^2}$. (b) Depicts shadows formed by massive particles (solid lines) and massless particles (dashed lines) for varying observer angles in the spacetime with $N=0$. (c) Exhibits shadows cast by massive particles (solid lines) and massless particles (dashed lines) for diverse values of the NUT parameter.}
    \label{fig:kerr_1}
\end{figure}

The shadows cast by massive particles with constant mass are depicted in Fig. \ref{fig:kerr_1}. In Fig. \ref{fig:kerr_1}a, the red line corresponds to the standard photon shadow with $\bar{m}_{E\infty}=0$. Unlike photons, the shadows of massive particles exhibit an increase in size as the particle energy decreases, i.e., as $\bar{m}_{E\infty}$ increases. Notably, the growth in shadow size from $\bar{m}_{E\infty}=0$ to 0.8 is comparable to the increase from 0.8 to 0.9, aligning well with the prediction from Eq. (\ref{eq:scale_approximation}), which yields $(R_{0.9}-R_0)/(R_{0.8}-R_0)\approx1.94$. The inset of Fig. \ref{fig:kerr_1}a presents the same set of shadows rescaled by a factor $\sqrt{1-\bar{m}_{E\infty}^2}$, revealing that the right side closely follows the approximation (\ref{eq:scale_approximation}), while the left side shows appreciable deviations. In Fig. \ref{fig:kerr_1}b, the evolution of the shadow for $\bar{m}_{E\infty}=0.9$ remains similar to the evolution of the photon shadow ($\bar{m}_{E\infty}=0$) as the observer changes its angle $\theta$. A notable feature of the shadows is the presence of fixed points at $X=0$ for different angles. Similarly, in Fig. \ref{fig:kerr_1}c the evolution of massive shadows for different $N$ mimics the progression of photon shadows. As $N$ increases, the shadow tends to become more circular, mitigating the effects of rotation since $a/\sqrt{M^2+N^2}$ decreases.

\begin{figure}
    \centering
    \includegraphics[width=1\linewidth]{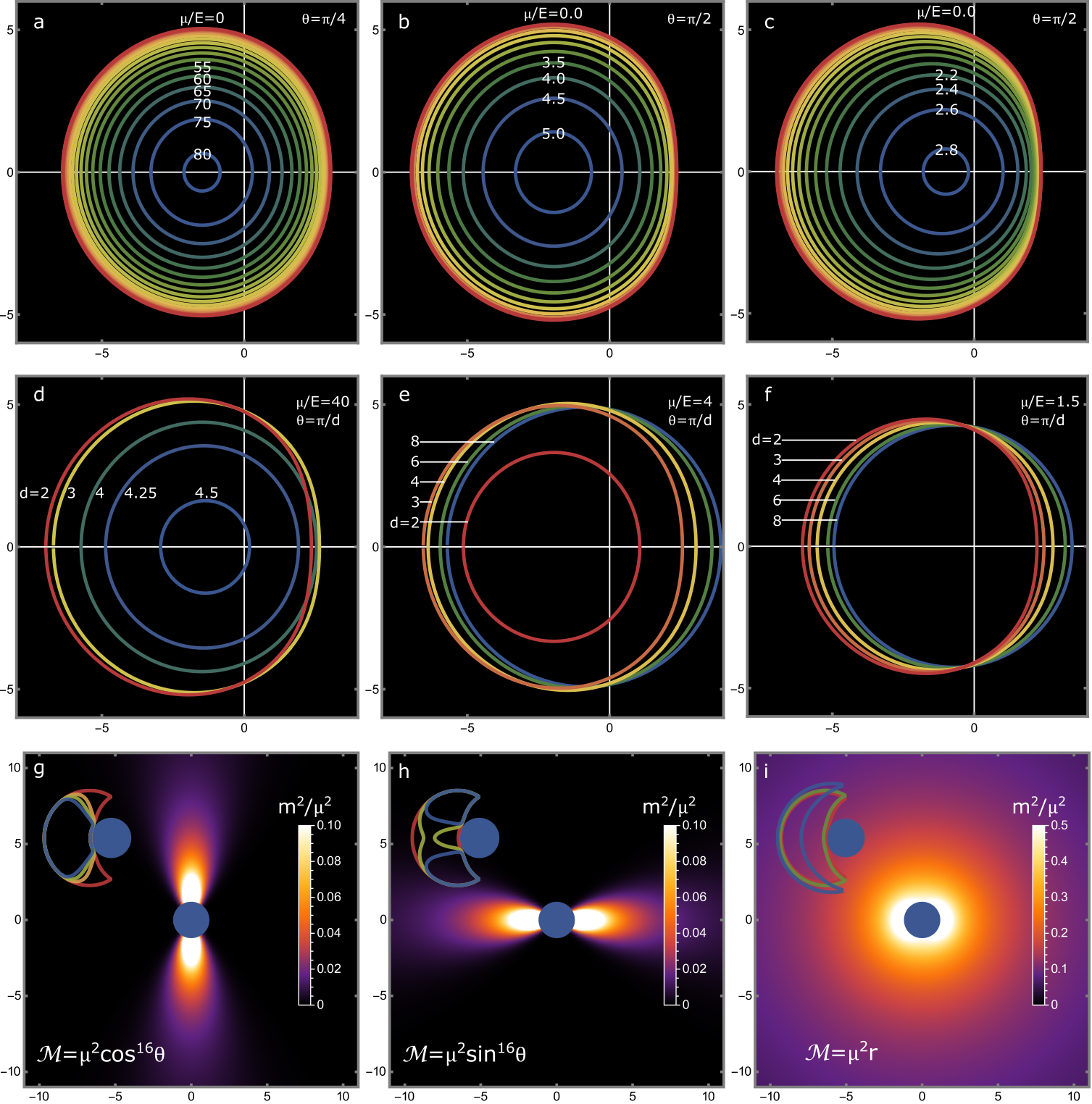}
    \caption{
    (a-c) Depiction of shadows formed by photons in plasma for varying values of the parameter $\mu$. (d-f) Illustration of shadows cast by photons in plasma at different observer angles, with a fixed parameter $\mu$ indicated in each respective panel. (g-i) Visualization of the distribution of the function $m_E^2/\mu^2$ in spacetime, where the blue disk represents the event horizon. The insets in panels (g-i) show the region of massive particles for various values of $\mu$, red color corresponds to the absence of plasma. Each column corresponds to a separate plasma distribution: (a,d,g) $\mathcal{M}=\mu^2 \cos^{16}\theta$, (b,e,h) $\mathcal{M}=\mu^2 \sin^{16}\theta$, (c,f,i) $\mathcal{M}=\mu^2 r$. The parameters are set to $M=1$, $a=0.98$, and $N=0$.}
    \label{fig:kerr_2}
\end{figure}
\begin{figure}
    \centering
    \includegraphics[width=1\linewidth]{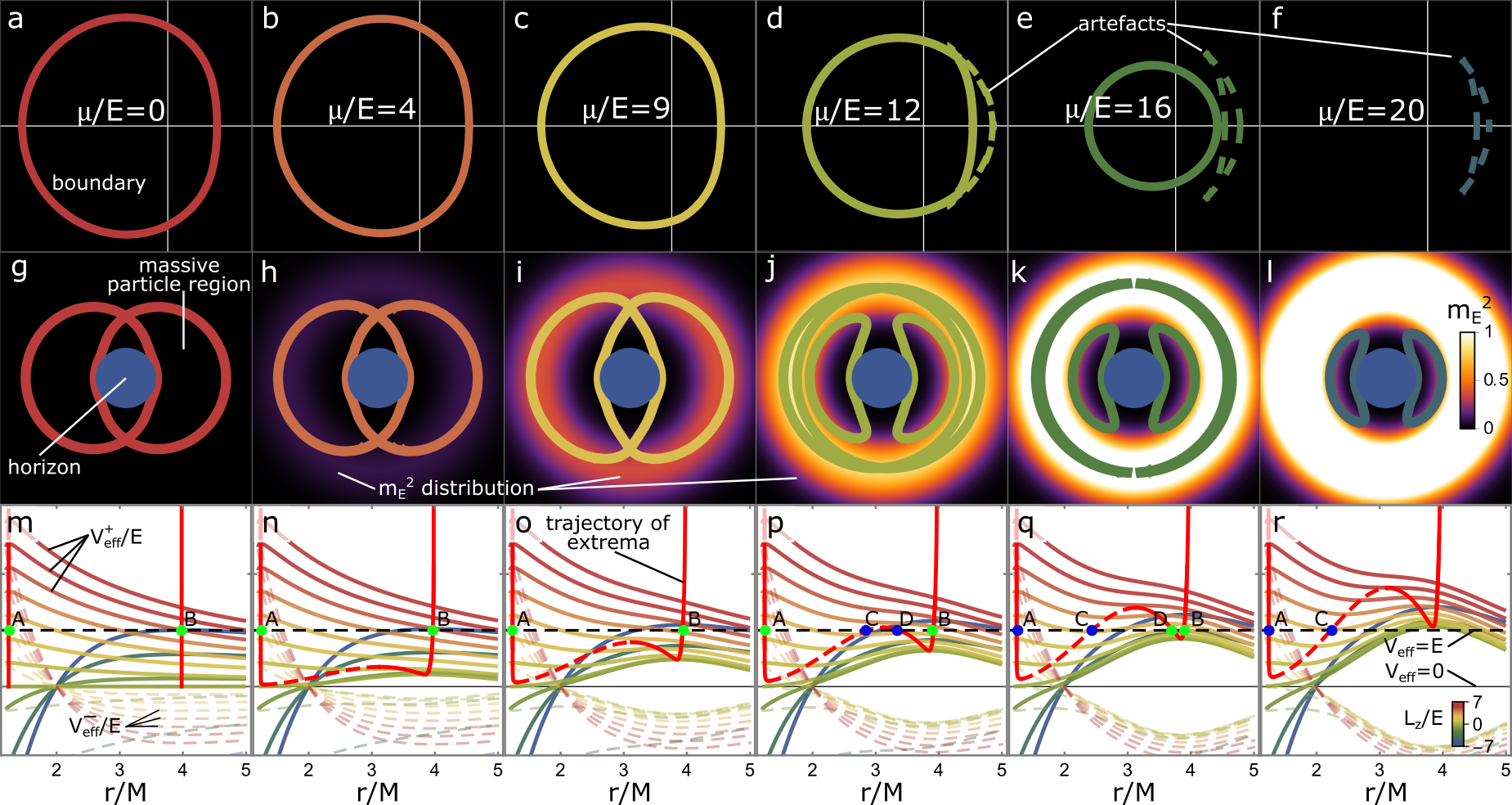}
    \caption{(a-f) Illustration of shadows formed by photons in the Kerr spacetime with plasma, considering a fixed $\mu/E$ at $\theta=\pi/2$. (g-l) Visualization of the massive particle region and the distribution of the function $m_E^2$ for a fixed $\mu/E$. (m-r) Presentation of the effective potential $V^\pm_\text{eff}/E$ at the equator for various values of $L_z/E$ with a fixed $\mu/E$. The red vivid line corresponds to points of maxima (solid) or minima (dashed) of the function $V^+_\text{eff}/E$. Green dots represent the shadow boundary, and blue dots indicate the artefact boundary. In all examples, parameters are set to $M=1$, $a=0.98$, $N=0$, and the plasma distribution is described by $\mathcal{M} = 10^{-5} \mu^2 r^8 \exp\{-(r - 2)^2/2\}$. Each column corresponds to a fixed value of $\mu/E$ as indicated in the figure.}
    \label{fig:kerr_3}
\end{figure}

To preserve the integrability of the dynamical system, we confine our attention to a specific class of plasma distributions, which gives rise to the following variable effective mass function \cite{Perlick:2023znh,Bezdekova:2022gib,Briozzo:2022mgg}
\begin{align}
    m^2=\frac{\mathcal{M}(r,\theta)}{r^2+(a \cos \theta + N)^2},\qquad
    \mathcal{M}(r,\theta) = \mathcal{M}_r(r) + \mathcal{M}_\theta(\theta), \qquad  \mathcal{M}^E=\mathcal{M}/\bar{E}^2,
\end{align}
which tends to zero in the distant regions away from the black hole. We exclusively examine the scenario where there is no gravimagnetic mass, $N=0$, focusing our attention on more plausible astrophysical systems. We will explore four distinct types of plasma distributions, as illustrated in Figs. \ref{fig:kerr_2} and \ref{fig:kerr_3}. In all instances, the plasma density is modulated by a multiplicative constant $\mu$.

The first type is characterized by the function $\mathcal{M} = \mu^2 \cos^{16}\theta$, reminiscing jets emanating from the black hole poles (Fig. \ref{fig:kerr_2}g). The shadow boundary, governed by Eq. (\ref{eq:BLXY}), depends on $\mathcal{M}$ and its derivative $\mathcal{M}'$ evaluated at the observer's angle. When the observer resides in the equatorial plane ($\theta=\pi/2$), both $\mathcal{M}$ and $\mathcal{M}'$ equal zero, rendering no effect of plasma on the shadow boundary. However, for observers at other positions, the presence of plasma becomes discernible. Shadows for various $\mu/E$ values with the observer at $\theta=\pi/4$ and shadows for different observer angles for $\mu/E=40$ are presented in Fig. \ref{fig:kerr_2}a and Fig. \ref{fig:kerr_2}d, respectively. In denser plasma with higher $\mu$ values, the shadow contracts and takes on a more circular shape. Additionally, with an increased observer altitude, the shadow exhibits further contraction. This behavior can be attributed to the prolonged interaction of photons with the plasma medium at higher altitudes, where the plasma exerts a focusing effect on photons, as discussed in Ref. \cite{Perlick:2017fio}. The inset of Fig. \ref{fig:kerr_2}g illustrates the impact of plasma on the massive particle region, revealing that the plasma, confined to the vicinity of the polar axis, primarily induces smoothing and shrinkage near the polar axes.

The second type of plasma, reminiscent of an accretion disk, is characterized by the function $\mathcal{M} = \mu^2 \sin^{16}\theta$ (Fig. \ref{fig:kerr_2}h). Analogous to the first type of plasma, for photons with lower energy (higher $\mu/E$), the shadow contracts and assumes a more circular shape (Fig. \ref{fig:kerr_2}b). However, in contrast to the first type, in this scenario, the shadow observed by an equatorial observer is smaller than that for non-equatorial observers (Fig. \ref{fig:kerr_2}e). The massive particle region associated with this plasma type may manifest topologically nontrivial features, as depicted in the inset of Fig. \ref{fig:kerr_2}h. This arises because photons may lack sufficient energy to penetrate the high-density plasma near the equator.

The first two types of plasma were characterized by a function $\mathcal{M}$ dependent solely on $\theta$. In contrast, the third type is described by an $r$-dependent function $\mathcal{M} = \mu^2 r$ reminiscent of nearly-spherical nebulae (Fig. \ref{fig:kerr_2}i). This scenario exhibits a similar focusing effect (Fig. \ref{fig:kerr_2}c), but the size of the shadow remains almost constant for observers at different altitudes (Fig. \ref{fig:kerr_2}f). This observation is not surprising, considering that the plasma distribution weakly depends on $\theta$ through the function $\Sigma$, specifically $m^2 = \mu^2 r / (r^2 + a^2 \cos^2\theta)$. The massive particle region is subtly displaced from the near-horizon region, where the plasma density is higher (inset in Fig. \ref{fig:kerr_2}i).

The fourth type of considered plasma is characterized by a shell-like distribution with the function $\mathcal{M} = 10^{-5} \mu^2 r^8 \exp\left\{-\frac{1}{2}(r - 2)^2\right\}$ (Fig. \ref{fig:kerr_3}g-l). To elucidate the dynamics of photons in plasma, one can construct the effective potential at the equator $\theta=\pi/2$:
\begin{equation}
    \dot{r}^2 = \frac{g^{tt}}{g^{rr}} E^2 \left(1 - V^+_\text{eff}/E\right)\left(1 - V^-_\text{eff}/E\right),
\end{equation}
where functions $V^\pm_\text{eff}/E$ are functions of the parameter $r$ parameterized by $L_z/E$ and $\mu/E$ (Fig. \ref{fig:kerr_3}m-r). The boundaries of the shadow correspond to the maxima of $V^+_\text{eff}/E$ located at the line $V^+_\text{eff}/E=1$ (green points in Fig. \ref{fig:kerr_3}m-r). For $\mu/E=0$, we have the standard boundary of the photon shadow (Fig. \ref{fig:kerr_3}a) with the massive particle region depicted in Fig. \ref{fig:kerr_3}g (in this case, it degenerates to the photon region).

The effective potential shown in Fig. \ref{fig:kerr_3}m has a maximum A (B) corresponding to the positive (negative) impact parameter $\rho_\phi=L_z/E$ and the right (left) point of the shadow boundary at $Y=0$. For $\mu/E = 4$ and $9$, the shadow boundary (Fig. \ref{fig:kerr_3}b,c), the massive particle region (Fig. \ref{fig:kerr_3}h,i), and the effective potential (Fig. \ref{fig:kerr_3}n,o) are almost the same, except for the following feature: there is a new maximum and a new minimum of $V^+_\text{eff}/E$ between points A and B. However, these new extrema do not play any role because they correspond to particles with lower energy than the energy fixed by the condition $\mu/E=4$ or $9$.

When $\mu/E$ is 12, there is one more maximum D and a new minimum C between the previous maxima A and B (Fig. \ref{fig:kerr_3}p). This corresponds to the appearance of two disjoint massive particle regions in Fig. \ref{fig:kerr_3}j. This leads to an artifact in the calculated shadow boundary, which should be considered unphysical. Indeed, the minimum should not play a role in the boundary formation, as it is hidden above the maximum of $V^+_\text{eff}/E$ and is not achievable by the geodesics connected with the distant observer. The new maximum is not related to the shadow boundary as well. If the energy is slightly smaller than the value of the maximum, the corresponding geodesics turn back from the massive particle surface. If the energy is slightly larger, then the corresponding geodesic will be turned back somewhere closer to the event horizon, as $V^+_\text{eff}$ achieves higher values near the horizon than at the local maximum D.

When $\mu/E$ is 16, the situation is different (Fig. \ref{fig:kerr_3}q). The points corresponding to the real shadow are D and B, but A and C correspond to the artifact (Fig. \ref{fig:kerr_3}e), though the massive particle region has the same structure (Fig. \ref{fig:kerr_3}k). Now, maximum A is lower than the outermost maximum of the same function $V^+_\text{eff}/E$, while the maximum D is a global maximum in the region outside the event horizon.

Finally, at $\mu/E=20$, there is no shadow at all (Fig. \ref{fig:kerr_3}f). The outermost massive particle region disappears (Fig. \ref{fig:kerr_3}l) with the corresponding maxima B and D (Fig. \ref{fig:kerr_3}r). At the same time, maximum A is not achievable, since there is another higher outermost maximum. Note that the transition of maxima between $\mu/E$ equal to 12 and 16 has its manifestation at the shadow. The right side of the shadow at $\mu/E=12$ is flat, but the artifact is quite circular, while for $\mu/E=16$ it is vice versa. The artifact is not physical in astrophysical applications. Nevertheless, it may find an application in quantum effects of condensed matter.

\subsubsection{Einstein-Maxwell-dilaton black holes}
The Einstein-Maxwell-dilaton model emerges from the 5D vacuum gravity through Kaluza-Klein dimensional reduction, wherein the dilaton constant is fixed at $\alpha=\sqrt{3}$. Consequently, the 5D gravity can be split into 4D gravity, an electromagnetic field $A_\mu$, and a dilaton field $\varphi$:
\begin{equation}
    ds_5^2 = e^{4\alpha\varphi/3} (d\chi - 2 A_\mu dx^\mu)^2 + e^{-2\alpha\varphi/3} \; ds_4^2,
\end{equation}
where $\chi$ represents the fifth compactified dimension. This model can be interpreted in two frameworks. In the first framework, 4D gravity and other fields are considered physical, and geodesics are calculated with respect to $ds_4$. In the second framework, the fields are regarded as auxiliary, but the 5D spacetime is considered physical, leading to geodesics calculated with respect to $ds_5$. Four-dimensional geodesics possess only a conformal Killing tensor, while 5D geodesics are fully integrable thanks to an exact Killing tensor. Nevertheless, 5D geodesic equations can be reduced to four dimensions (see App. B in Ref. \cite{Bogush:2020obx}). Such 5D geodesics, when reduced to 4D, obey the usual 4D geodesic equations, with the only difference being a variable effective mass, similar to Eq. (\ref{eq:geodesic_equation_var_m}). The momentum along the fifth dimension generates electric charge, which we set to zero, as we do not consider charged particles in this paper. Therefore, the effective mass is:
\begin{equation}
    m^2_\text{eff} = m^2 e^{-2\alpha\varphi/3}.
\end{equation}
Given that the dynamical system is reduced and properly truncated from another system possessing an exact Killing tensor, the integrability is inherited from 5D to 4D, as shown below. As an illustrative example, we will employ the solution discovered in Ref. \cite{Rasheed:1995zv}:
\begin{equation}
    \Delta = r(r-2M) - 3D^2 + Q^2 + P^2 + a^2,\qquad
    \Sigma = \sqrt{AB},\qquad
    e^{-2\alpha\varphi/3} = \sqrt{A/B},
\end{equation}
where the functions $A$ and $B$ take the form of separable quadratic polynomials in terms of $r$ and $\cos\theta$. These functions, along with $\omega$, are detailed in Ref. \cite{Rasheed:1995zv}, and they involve five parameters: mass $M$, rotation parameter $a$, electric and magnetic charges $Q$ and $P$, and dilaton charge $D$ (note that the dilaton charge here differs from the scalar charge $\Sigma$ in Ref. \cite{Rasheed:1995zv} by a factor of $\sqrt{3}$). The charges are constrained by the equation:
\begin{equation}
      \frac{Q^2}{D + M}
    + \frac{P^2}{D - M} = 2D,
\end{equation}
which has three solutions for $D$, but it is established in Ref. \cite{Bogush:2020obx} that only one of them can appropriately represent a black hole solution. The effective mass is:
\begin{equation}
    m^2_\text{eff} = m^2 \sqrt{\frac{A}{B}},
\end{equation}
which satisfies the integrability condition, i.e., $\Sigma m^2_\text{eff} = m^2 A$ is separable.
\begin{figure}
    \centering
    \includegraphics[width=1\linewidth]{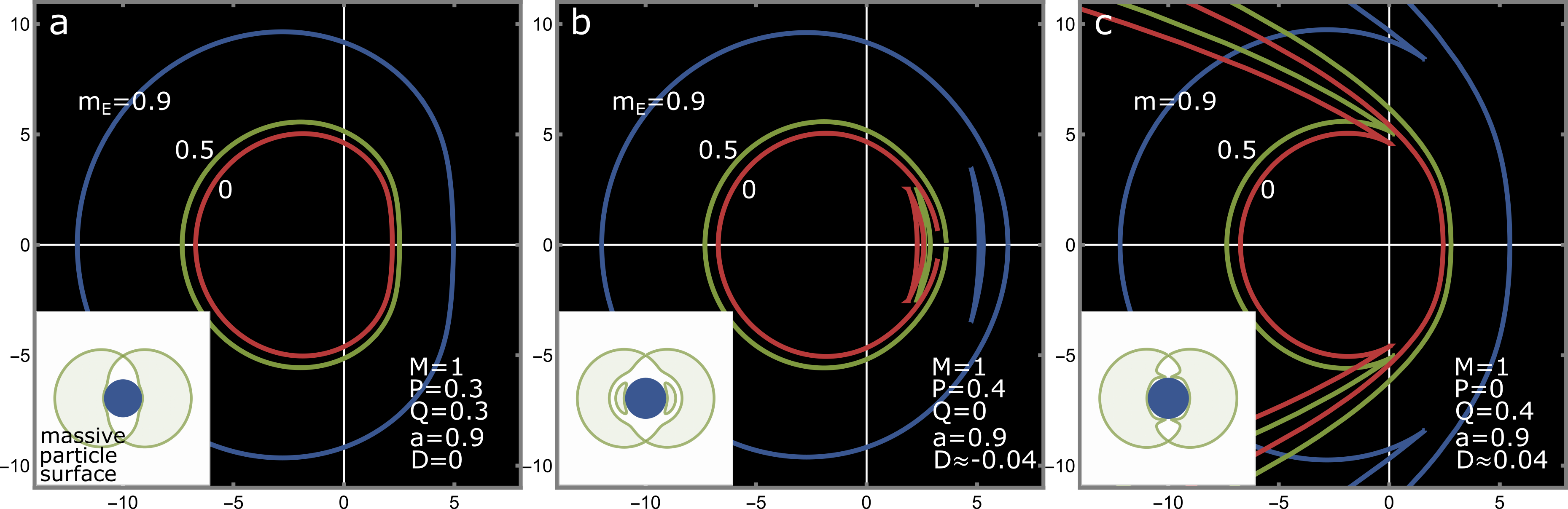}
    \caption{
    Shadows formed by massive particles in the EMD black hole spacetime for different $m_E$. Each panel corresponds to specific solution parameters indicated within. Insets feature images of the massive particle surface, where blue disks denote the event horizon.}
    \label{fig:emd}
\end{figure}

We explore three examples (Fig. \ref{fig:emd}). In all instances, the deviation of the effective mass, $(m_\text{eff}^2 - m^2)/m^2$, does not exceed $\pm10\%$. The first case involves equal electric and magnetic charges, and zero dilaton charge (Fig. \ref{fig:emd}a), the second is magnetically charged (Fig. \ref{fig:emd}b), and the third one is electrically charged (Fig. \ref{fig:emd}c). The first example exhibits a regular shadow boundary. In the second case, a crescent-shaped artifact, as discussed in the Kerr example, is present. The third example features a ``fishtail'' type of artifacts, which has been discussed in Ref. \cite{Perlick:2023znh} for photons in plasma in the Kerr spacetime. This type of artifact shares similar roots with the crescent artifact, i.e., the existence of unachievable minima/maxima of the effective potential. However, in our case, the ``fishtail'' appears without plasma, arising from the interaction with the dilaton field or the existence of the fifth compactified dimension, depending on the model's interpretation. More examples of shadows cast by photon in plasma in EMD model are given in Ref. \cite{Badia:2022phg}. Numerous other gravitational models with different plasma distributions have been reviewed in Refs. \cite{Bisnovatyi-Kogan:2017kii,Abdujabbarov:2016efm,Fathi:2021mjc,Kumar:2023wfp,Bezdekova:2022gib,Perlick:2023znh,Frost:2023enn,Alloqulov:2023ahx,Chen:2022scf,Kala:2022uog,Cunha:2018acu,Ditta:2023lny}.

\section{Conclusions}
The goal of this article was to obtain, for a general spacetime with a slice-reducible exact or conformal Killing tensor, explicitly coordinate-independent analytical expressions defining the photon/massive particle regions (\ref{eq:MPR}) and the contours of gravitational shadows (\ref{eq:Theta_Phi}),  (\ref{eq:XY_general_all}) for neutral particles with variable mass.
This framework is directly applicable to neutral elementary and composite  particles, such as neutrinos, photons, non-ionized atoms, etc. Using the concept of coordinate-dependent mass it naturally expands to include the important case of photons in nonmagnetized pressureless plasma whose distribution inherits symmetries of spacetime. General conditions on the mass function are formulated in coordinate-independent way which ensure the integrability of the equations of motion in integrable spacetime.
In the absence of electric/magnetic charges of black holes and magnetic fields in their vicinity, the electric (magnetic) charge of moving particles is irrelevant, effectively expanding the scope to the case of electrons, protons or alpha particles. The concept of massive particle surfaces can be generalized to charged particles in  presence of  electromagnetic fields possessing the same symmetries, so our framework can be extended in the future to include charged particles more generally.

The presented expressions have the advantage of a completely invariant construction, combining simplicity and universality. Notably, there is no need to explicitly use  coordinates ensuring separation of variables such as Boyer-Lindquist coordinates. However, the latter simplifies the calculations in many cases. In fact, the entire structure of the gravitational shadow and the   massive particles region is determined by a single shadow matrix (\ref{eq:shadow_matrix}), which has a surprisingly simple form and is deeply connected to the massive particle surfaces. We hope that this framework will contribute to analytical studies of gravitational shadows and integrable systems, especially in determining general constraints on shadow size and other observable quantities.

We also examine the asymptotic behavior of the results and derive shadow formulas for an asymptotically distant observer (\ref{eq:XY_infty}). It is obvious that the current experimental capabilities of radio and neutrino astronomy are insufficient to make deep conclusions about the details of the dynamics of elementary particles in curved spacetime and their deviations from the Kerr picture, since this requires much greater resolution. Although we hope that the required resolution will be achievable in the future, the presented structure is promising also for the analysis of analog models of gravity or the description of (quasi-)particles in crystals. In these scenarios, either the effective mass is variable, or the crystallographic defects can be effectively described in terms of differential geometry, or both.

Using separation coordinates, the result simplifies to a shadow formula for the general Benenti-Francaviglia metric (\ref{eq:XY_BF}) and (\ref{eq:XY_asymptotic}) (including the conformal generalization) and a general metric allowing Boyer-Lindquist coordinates (\ref{eq:BLXY_G}). The use of these formulas facilitates the direct construction of shadow images of massive particles and photons in plasma by simply replacing the explicit components of the metric in its original form without the need for auxiliary calculations. This versatility extends to a wide range of solutions in supergravity and the low-energy limits of string theory.

The developed scheme is illustrated with various examples in spacetime of Kerr-NUT and EMD black holes, which successfully reproduce the focusing effect of a plasma medium on low-energy photons. We explore the effects of plasma distributions resembling jets, accretion disks, and near-spherical nebulae, discussing their impact on the massive particle region. In addition, we provide an example of 5D geodesics reduced to a 4D system with a variable effective mass, which can be interpreted as interaction with a scalar field in the EMD model. The framework reproduces two types of edge artifacts -- one crescent-shaped and the other known as a fishtail. We found that the flattened side of the shadow can be accurately approximated by a scale factor of $(1-\bar{m}^2_{E\infty})^{-1/2}$, while the rounder part exhibits visible deviations described by the mass-dependent term of the shadow matrix $\mathcal{S}^{ab}$.
 
We hope that the developed framework will contribute to the understanding of the general patterns of the formation of shadows cast by both massive and massless particles, both in vacuum and plasma environments. This concept may find applications in astrophysics, analog models of gravity, and condensed matter physics.

\begin{acknowledgments}
This work was supported by Russian Science Foundation under Contract No. 23-22-00424.
\end{acknowledgments}

\bibliography{main}

\end{document}